\documentclass{tADP2e}

\newcommand{\beq}{\begin{equation}}
\newcommand{\eeq}{\end{equation}}
\newcommand{\beqa}{\begin{eqnarray}}
\newcommand{\eeqa}{\end{eqnarray}}
\renewcommand{\a}{\alpha}
\newcommand{\abs}[1]{\vert#1\vert}
\renewcommand{\b}{\beta}
\renewcommand{\d}{{\rm d}}
\newcommand{\del}{\delta}

\newcommand{\eps}{\varepsilon}
\newcommand{\g}{\gamma}

\newcommand{\frat}[2]{{\tfrac{#1}{#2}}}%{{\textstyle{#1\over#2}}}
\newcommand{\hb}{\overline{h}}

\newcommand{\nub}{\overline{\nu}}
\newcommand{\mean}[1]{\langle#1\rangle}
\renewcommand{\o}{\omega}
\newcommand{\s}{\sigma}

\renewcommand{\O}{\Omega}

\newcommand{\xid}{{\xi_{\rm d}}}
\newcommand{\xis}{{\xi_{\rm s}}}
\newcommand{\mud}{{\mu_{\rm d}}}
\newcommand{\mus}{{\mu_{\rm s}}}

\def\half{\tfrac{1}{2}}

\newcommand{\e}{\mathrm{e}}
\newcommand{\etal}{{\it et al\/}\ }

\usepackage{color}
\newcounter{notecounter}

\begin{document}

% \jvol{00} \jnum{00} \jyear{2015} \jmonth{January}

\title{Storing and retrieving long-term  memories: cooperation and competition in synaptic dynamics}
\author{%
  \name{Anita Mehta${}^\ast$\thanks{${}^\ast$Email: anita@bioinf.uni-leipzig.de}}%
  \affil{Dipartimento di Fisica, Universit\`a di Roma La Sapienza, P.~A.~Moro 2, 00185 Roma, Italy}
                                                     and
  \affil{Institut f\"ur Informatik, IZBI, Universit\"at Leipzig, H\"artelstrasse 16--18, 04107 Leipzig, Germany}%
}

\maketitle

\begin{abstract}
  We first review traditional approaches to memory storage and formation, drawing on the literature
  of quantitative neuroscience as well as statistical physics. These have generally focused on the fast dynamics of neurons;
  however, there is now an increasing emphasis on the slow dynamics of synapses, whose weight changes are held to be responsible
  for memory storage. An important first step in this direction was taken in the context of Fusi's cascade model, where complex synaptic
  architectures were invoked, in particular, to store long-term memories. No explicit synaptic dynamics were, however, invoked in that work.
  These were recently incorporated theoretically using the techniques used in agent-based modelling, and subsequently, models of competing and cooperating synapses were formulated.
  It was found that the key to the storage of long-term memories lay in the {\it competitive} dynamics of synapses. In this review, we focus on models of synaptic competition and cooperation,
  and look at the outstanding challenges that remain.  
\end{abstract}

\begin{classcode}
87.18.Sn Neural networks and synaptic communication, 87.19.lv Learning and memory, 89.75.Da Systems obeying scaling laws, 05.40.-a Fluctuation phenomena, random processes, noise, and Brownian motion
\end{classcode}

\begin{keywords}
Synaptic plasticity; competitive learning; power-law forgetting; competitive synaptic dynamics; long-term memory
\end{keywords}

\section{Introduction}\label{intro}

Memory~\cite{kandel1, kandel2} and its mechanisms have always attracted a great deal of interest~\cite{ebb}.
It is well known that memory is not a monolithic
construct, and that memory subsystems corresponding to episodic, semantic or
working memory exist~\cite{squire}.
We focus here on explicit memory, which is the memory for events and facts.

Models of memory have, themselves, long been studied in the field of mathematical psychology: the article by Raaijmakers and Shiffrin~\cite{raaijmakers2002models} provides a valuable review of models that existed well before the neural network models with which most physicists are familiar, began to appear. 
Here, memory was assumed to be distributed over a large set of nodes and an item was defined by  the pattern of activation over a set of nodes.  This was propagated through a
network of links whose geometry and weights determined the output. Such models of storage and retrieval are discussed at length in~\cite{raaijmakers2002models}, but in the interests of a historical presentation, we briefly
describe the earliest example known as 
the `brain state in a box' model, or BSB~\cite{anderson1977distinctive}. In this model, items are vectors while learning is  represented by changes in synaptic strengths. For any
such  pair of items, the synaptic strengths between the input and output layers are
modified in such a way that considerable storage and retrieval is possible, even in the presence of noise. There have in parallel been a lot of suggestions regarding the 
way in which working memory actually functions: from the point of view of the current review, the most important distinction between these is that forgetting involves temporal decay in the research of Baddeley 
and co-workers~\cite{baddeley2003working,baddeley1974ga} and that it does not, in the work of Nairne and co-workers~\cite{nairne2002remembering,neath2012arguments}.
 Although a detailed discussion of these psychological (and somewhat empirical) models is beyond the scope of this review, they do
indeed offer fertile ground for mathematical modellers who would wish to construct quantitative models of working memory.

 In general, memories are acquired by the process of
learning. Simply put,
patterns of neural activity change the strength of synaptic connections within the brain, and the reactivation of these constitutes memory~\cite{martinmorris}.
In this context, we first review
the different kinds of learning to which a network can be subjected~\cite{simpson2009theoretical}.
These are respectively: supervised, reinforcement, and unsupervised learning. In {\it supervised
learning}, the goal is to learn a mapping between given input and output
vectors, as, for instance, when we classify the identity of items in a list.
In {\it reinforcement learning}, the goal is to learn a mapping between a set of
inputs or actions in a particular environment and 
some measure of reward.
In {\it unsupervised learning}, the network is provided with no feedback at all.
Rather, synaptic strength changes occur according to a learning rule based
only on pre- and post-synaptic activity, with no reference to any desired output.
The pattern of synaptic strengths that results in this case, depends on the nature of the learning rule and
the statistical structure of the inputs presented. It is this kind of learning with which this review will be chiefly concerned.

The somewhat bland statement above, of memories being acquired by a process of learning, actually  pushes a lot of puzzles under the rug. Why is it that some memories are quickly forgotten, while others last a lifetime?
One hypothesis is that important memories are transferred, via
their synaptic strengths, to different parts of the brain that are less exposed to ambient noise.  In particular, during a process known as {\it synaptic consolidation}~\cite{meta3}{\footnote{In the literature, this is sometimes referred to as systems consolidation, while synaptic consolidation is traditionally used to describe the molecular mechanism that leads to the maintenance of synaptic plasticity.}}, memories
that are first stored in the hippocampus are transferred to other
areas of the cortex~\cite{kirwan, smith}; this transfer can happen while the events are rerun during sleep~\cite{diba}.
The case of the famous patient HM~\cite{HM} whose hippocampus
was removed following epilepsy 
reinforces this hypothesis:
HM retained old memories from before his surgery, but he
could barely acquire any new long-term memories.

There is yet another mechanism for memory consolidation which happens at the synaptic level, involving the mechanism of {\it synaptic plasticity}, whereby synapses change their strength.
Short-term
plasticity (STP) occurs when the change lasts up to a few minutes, 
while long-lasting increases/decreases of synaptic strength are known respectively as long-term potentiation/depression (LTP/LTD); LTP was first discovered experimentally by
Bliss and Lomo~\cite{blisslomo} in 1973.
Long-term plasticity is further subdivided into
early-long-term plasticity  (e-LTP) when synaptic changes last
 up to a few hours and
late-long-term plasticity (l-LTP), when they last from beyond typical experimental durations of 10 hours to possibly a lifetime.
Such late-long-term plasticity
also falls within the terminology of synaptic consolidation~\cite{clopath}; here, relevant memories are
consolidated {\it within} the synapses concerned, so that new memories
can no longer alter previously consolidated ones. The two most important theoretical models of this second kind of synaptic consolidation involve a process called
{\it synaptic tagging}~\cite{clopath1, barrett, clopath}. The hypothesis is that a single,
brief  burst of high-frequency stimulation is enough to induce e-LTP, and its expression does not require protein synthesis.
On the other hand, l-LTP  can be induced
by repeated  bursts of high-frequency stimulation, which leads to an increase in synaptic strength until saturation is reached.  There is also a view~\cite{barrett} that more stimulation does not increase the amount of synaptic weight change at individual synapses, but rather increases the duration of weight enhancement.  In this case, it has been shown
that protein synthesis is triggered at the time of induction. Also, it was found that e-LTP at one synapse could be converted to l-LTP
if repeated bursts of high-frequency stimulation are given to other inputs of the same neuron during a short period before or after the induction of e-LTP at the first synapse~\cite{freymorris}. This discovery led to the hypothesis that such stimulation initiates the creation of a `synaptic tag' at the
stimulated synapse, which is thought to be able to capture plasticity-related proteins. The general
framework for these hetero-synaptic effects is called {\it synaptic
tagging and capture}, for the details of which
 the  reader is referred to~\cite{clopath, barrett}. 
 
 It should be mentioned here that because of the interdisciplinary nature of the field, much of the discussion in the literature~\cite{graupner, miyamoto} involves
 terminology such as 'plasticity induction and maintenance', to refer respectively to short-term and long-term plasticity changes. Specifically, in~\cite{miyamoto}, the author's findings reinforce
 the intuition  that  LTP induction and maintenance would lead respectively to short- and long-term memory. Thus in the following, models manifesting short-term memory involve only plasticity induction,
 while plasticity maintenance is responsible for the manifestation of long-term memory in the models that form the core of this review.

  Finally, some of the most recent developments in the modelling of memory acquisition and maintenance involve the concept of engrams~\cite{kitamura}; here, memories may be reconstructed by single neuronal activation.The underlying idea is that a big network of neurons is involved in memory acquisition, with several connections being modified; these may be lost over time or in an activity-dependent manner such that memory is virtually supported by a single connection, and later  reconstructed. This mechanism suggests that memory reactivation may not rely on the same network involved in its acquisition, but rather on the reconnection of neurons that may have similar responses. The authors of~\cite{kitamura} also suggest that  memories at the time of acquisition are already stored in the cortex, instead of being transferred from the hippocampus to the cortex as suggested in~\cite{kirwan, smith}.

To sum up: memory formation is 
 a complicated phenomenon related to neural activities,
brain network structure, synaptic plasticity~\cite{vanrossum} and synaptic consolidation~\cite{clopath}. 

We will provide an overview of some of the more traditional approaches, involving neural networks -- both those based on detailed biophysical principles, and those that
were explored by statistical physicists starting from the seminal work of Hopfield~\cite{hopfield1982}. Much of this has already been extensively reviewed, so the focus of the present review
comprises questions like: how can short-term and long-term memory coexist in our brains? While it is known that short-term memory is ubiquitous, what are the synaptic mechanisms needed for
long-term memory storage?

It is well known that too much plasticity causes the erasure 
of old memories, while too little plasticity does not allow for the quick storage of new memories. This {\it palimpsest paradox}~\cite{nadal, parisi}
has been at the heart of the quandary faced by modellers of synaptic dynamics.
While synaptic consolidation does indeed provide some insights into this, neuroscientists~\cite{turr2,avy}
have typically focused on synaptic plasticity~\cite{takeu}, for which
increasingly sophisticated models
have emerged over the years~\cite{book1,book2,book3}. There 
 are two broad classes:
{\it biophysical} models, which incorporate details at the molecular level, and
{\it phenomenological} models, which relate neuronal activity to synaptic
plasticity. 
It is the latter class of models that we will focus on in this 
review, both because they are more amenable to statistical physical
techniques and because they account for higher-level phenomena like memory
formation.
 Such modelling, while it may not include details of specificities involving chemical and biological processes in the brain, can
outline possible mechanisms that take place in simplified structures. For example, the study of neural networks~\cite{book1,book2,book3}, while it greatly
 simplifies biological structures in order to make them tractable,
has still been able to make an impact on the parent field. In particular, neural networks such as the Hopfield model~\cite{hopfield1982,hopfield1984} have been 
extensively investigated via methods borrowed from the statistical
physics of disordered and complex systems~\cite{ags1,ags2,ags3}. In these models, memories are stored as patterns of neural activities,
 which correspond both to low-energy states and to attractors of the stochastic dynamics of
the model. 

What this class of phenomenological models lacks in biological detail, it typically makes up for in
 minimalism. Abbott, one of the pioneers in this field, summed up its virtues thus~\cite{abbott2008theoretical}:
\begin{quote}
Identifying the minimum set of features needed to account for a particular phenomenon and describing these accurately enough to do the job is a key component of model building. Anything
more than this minimum set makes the model harder to understand and more difficult to evaluate. The term `realistic' model is a sociological rather than a scientific term. The truly realistic
model is as impossible and useless a concept as Borges' map of the empire that was of the same scale as the empire and that coincided with it point for point.
\end{quote}

Within this class of models, there is yet another divide; there are models which focus on the fast dynamics of neurons, and then those that focus on the slow dynamics of synapses.
We will review each one in turn. In particular, in the second case, we will focus on the nature of synaptic dynamics, which involve {\it competition} and {\it cooperation}~\cite{miller}. 
There is abundant evidence that correlation-based rules of synaptic cooperation, which lead to the outcome `neurons that fire together, wire together', are followed in many organisms; the latter is known as Hebb's rule, due to the pioneering work of Hebb in establishing it~\cite{hebb}. In synaptic cooperation therefore, synapses that work together are rewarded by being strengthened.  However, synapses also have a competitive side: while some synapses grow stronger and prosper, others, which left to themselves would also have strengthened, instead weaken. (An example of this can be seen in the process of ocular dominance segregation~\cite{miller1989ocular}, where competitive correlations ensure that inputs to the left and right eye, though they fire together, do {\it not} wire together). Of these two processes, synaptic cooperation is by far the more commonly used in mathematical modelling; however, its unbridled prevalence leads to instabilities, for which
synaptic competition provides a cure. From a more biological standpoint, synaptic competition is a concept that has long found favour
with the neuroscience community~\cite{avy}; however, its use is relatively recent in the context of statistical physics models. The present review accordingly emphasises those approaches
where synaptic cooperation and competition are key.

We begin this article with a review of the Hopfield model (Section~\ref{hopf}), where we describe the model as well as its use in storing and retrieving random patterns.
We then turn to phenomenological models of synaptic plasticity (Section~\ref{phen}), which are further classified as {\it rate-based models} (Section~\ref{rate}) and
 {\it spike-time-dependent plasticity models} (Section~\ref{stdp}),
where the synaptic strength is always treated as a {\it continuous} variable. A change of key sets in in Section~\ref{state}, when synapses are discretised, with the further possibility~(Section~\ref{casc}) of occupying a multiplicity of states. In the following section (Section~\ref{comp}), we present an extensive review of the neuroscience literature to do with the perceived need for synaptic competition. These ideas are implemented in 
Section~\ref{statsyn} 
where, in particular, synaptic strengths are discretised and competitive dynamics embedded, using  
 tools from statistical physics. 
In the Discussion (Section~\ref{disc}), we summarise the state of the literature, and discuss some future challenges.

\section{The Hopfield model}\label{hopf}

Appropriately for the readership of this journal, we start by reviewing the Hopfield model, both because this is one of the seminal contributions
of physics to the field, and also because it is the basis on which a large class of models (STDP, cf. Section~\ref{stdp}) is based.

In 1982, John Hopfield introduced an artificial neural
network to store and retrieve memory like the human brain~\cite{hopfield1982,hopfield1984}. In such a fully connected network of $N$ neurons, 
there is a connectivity (synaptic) weight $J_{ij}$ between 
any two neurons $i$ and $j$, 
which is symmetric
so that $J_{ij} = J_{ji}$, and $J_{ii} = 0$.
% The state of a neuron $i$ is given by $+1$ when it is firing, and $-1$ when it is not. This state is determined by a local field 
%$h_i =\frac{1}{N-1}\sum_{j\ne i} J_{ij}x_j$, where  the $x_j$ are the inputs of the neurons $j \neq i$.
 Such a network
is initially trained to store a number of patterns or memories.
It is then able to recognise any of the learned patterns by
exposure to only partial or even some corrupted
information about that pattern, i.e., it eventually settles
down and returns the closest pattern or the best guess.

We present here a simple picture of memory storage and retrieval along the lines of~\cite{abbott1990learning}.
Each neuron is characterised
by a variable $S$ which takes the value $+1$ if the neuron is firing and $-1$ if the neuron
is not firing. 
At time $t + 1$ the neuron labelled by the index~$i$, where $i = 1,2,3,{\dots},N$ for a system of $N$ cells,
fires or does not fire based on whether the total signal it is receiving from other cells
to which it is synaptically connected is positive or negative. Thus, the basic dynamical
rule is
\begin{equation}\label{eq:abbot1.1}
  S_i(t+1) = \operatorname{sgn} \Biggl(\sum_{j=1}^NJ_{ij}S_j(t)\Biggr),
\end{equation}
where $J_{ij}$ is a {\it continuous} variable representing the strength of the synapse connecting cell $j$ to cell $i$. 
The basis of a network associative memory  is that the above dynamics
can map an initial state of firing and non-firing neurons, $S_i(0)$, to a fixed pattern, $\xi_i$,
which remains invariant under it.
Various memory patterns $\xi_i^\mu$ %\inlinemath{$xi_i^mu$} 
for
$\mu = 1,2,3,{\dots}, P$ which
do not
change under the above transformation act as fixed-point attractors; initial inputs
$S_i(0)$ are mapped to an associated memory pattern $\xi_i^\mu$ %\inlinemath{$\xi_i^mu$}
if the overlap
$\sum\xi_i^\mu S_i(0)/N$ %\inlinemath{$Sigma../N$}
is
close enough to one. How close this overlap must be to one, or equivalently how well
the initial pattern must match the memory pattern in order to be mapped to it and
thus associated with it, is determined by the radius of the domain of attraction of the
fixed point.
The issue of domains of attraction associated with a fixed point has never been
completely resolved. The sum of all synaptic inputs at site $i$,
\begin{equation}\label{eq:abbot1.2}
  h_i^\mu = \sum_{j=1}^NJ_{ij}\xi_j^\mu,
\end{equation}
known as the local field, is the signal which tells cell $i$ whether or not to fire when
$S_j = \xi_j^\mu$ %\inlinemath{$Sj = xi_j^mu$} 
for all $j \neq i$.
In order for a memory pattern to be a stable fixed point of the
dynamics,  the local field must have the same sign as 
$\xi_i^\mu$ %\inlinemath{$xi_i^mu$} 
or equivalently
\begin{equation}\label{eq:abbot1.3}
  h_i^\mu\xi_i^\mu>0\:.
\end{equation}
We will call the quantities $h_i^\mu\xi_i^\mu$ %\inlinemath{ hi...}
the aligned local fields. It seems reasonable to assume
that the larger the aligned local fields are for a given $\mu$  value the stronger the attraction
of the corresponding fixed point 
$\xi_i^\mu$ %\inlinemath{$xi_i^mu$} 
and so the larger its domain of attraction. This
reasoning is almost right, but it leaves out an important feature of the above dynamics.
Multiplying $J_{ij}$ by any constants  has absolutely no effect since the
dynamics depends only on the sign and not on the magnitude of the quantity
$\sum J_{ij}S_j$. %\inlinemath{$ SigmaJ,,.S,$}
Since the quantities
$h_i^\mu\xi_i^\mu$ %\inlinemath{ h:(r)}
change under this multiplication they alone cannot determine
the size of the basin of attraction. Instead, 
it has been found that
quantities known as stability parameters and given by
\begin{equation}\label{eq:abbot1.4}
  \gamma_i^\mu=\frac{h_i^\mu\xi_i^\mu}{\vert J\vert_i},
\end{equation}
where we define
\begin{equation}\label{eq:abbot1.5}
  \vert J\vert_i=\Biggl(\sum_{j=1}^NJ_{ij}^2\Biggr)^{1/2},
\end{equation}
provide an important indicator of the size of the basin of attraction associated with the
fixed point 
$\xi_i^\mu$. %\inlinemath{$xi_i^mu$}
Roughly speaking, the larger the values of the 
$\gamma_i^\mu$ %\inlinemath{$gamma_i^mu$}
the larger the domain
of attraction of the associated memory pattern. In order to construct an associative memory one must find a matrix of synaptic
strengths $J_{ij}$ which satisfies the condition of stability of the memory fixed points 
and has a specified distribution of values for the
$\gamma_i^\mu$ %\inlinemath{$gamma_i^mu$}
giving the domain of attraction
which is desired.

Notice that in the above, the synapses are used for storage and retrieval of memories, as well as a way of updating the neuronal states; in other words,
they are not explicitly updated.  

\section{Phenomenological models of synaptic plasticity}\label{phen}

We move on now to models where plasticity is invoked, i.e., where synapses are explicitly updated.
The assumption here is that neuronal
firing rates are, in their turn, responsible for synaptic strengthening or weakening. The basic principle at work is Hebb's rule~\cite{hebb}, which as mentioned above, says that `cells that fire together, wire together'. Another way of viewing this rule is to say that simultaneous events over a period of time suggest a causal link, and many {\it rate-based} models of synaptic plasticity have been formulated on this basis.
However, and more recently, a great deal of attention has been paid to a much stricter definition of causality via the field of 
{\it spike-timing-dependent plasticity} (STDP)~\cite{markram1995action,gerstner1996neuronal}:
here,
 synaptic strengthening only occurs if one of the neurons is 
systematically active just before another one. 
In addition to realising the Hebbian condition that
 a synapse should
be strengthened only if it
constitutes a causal link between the firing of pre- and post-synaptic neurons,
STDP also
leads to the {\it weakening} of synapses
which connect neurons whose firings are temporally correlated, but where the firing is {\it not} causally ordered.
 
We briefly review these two classes of models below.

\subsection{Rate-based models}\label{rate}

Here,
the rate of pre- and post-synaptic activities measured over some time period
determines the sign and magnitude of synaptic plasticity.
The activities are
modelled as continuous variables, corresponding to a suitable average of neuronal  firing rates.  The rate of change of synaptic strength or weight $J_i$  at synapse $i$ is 
modelled as a function of the pre-synaptic input $x_i$ at that synapse, the post-synaptic output activity
$y$, the weight itself, and, in the most general case, the weights of other synapses: 
\beq
\frac{\d J_i}{\d t} = f(x_i,y, J_i,J_j).
\label{dwdt}
\eeq

Without the {\it competition} from other synapses $J_j$,
synaptic weights could grow uncontrollably. Before the explicit inclusion of synaptic competition,
this instability was combated in two ways; in
Oja's~\cite{oja} model,  Hebbian plasticity was augmented with a decay term, so that weights equilibrated to the first principal component of the
input correlation matrix. Another way forward was shown by the BCM model~\cite{bcm} which explicitly included both LTP and LTD regions, with a
sliding threshold separating them; when synaptic weights became too large, the threshold shifted so that any further activation led to synaptic depression.
Subsequently, indirect ways of including synaptic competition (Section~\ref{comp}), such as the normalisation of the total synaptic weights, were included in the modelling; more recently,
there have been a number of approaches where synaptic weights are discretised (Section~\ref{state}) and competition explicitly implemented (Section~\ref{statsyn}).

\subsection{Models of spike-timing-dependent-plasticity}\label{stdp}

Spike-timing-dependent-plasticity (STDP)  provides the answer to the following question:  For neurons embedded in a network which are bombarded with millions of inputs, which ones are important? Which information should a given neuron `listen' to, and pass along to downstream neurons? These are the formidable questions that the vast majority of neurons in the brain have
to solve during brain development and  learning. The crucial link is causality -- if one of the cells is active systematically just slightly before another, the firing of the first one might have a causal link to the firing of the second one and this causal link could be remembered by increasing the wiring of connections. Theoreticians in the mid-1990's realized
just how important temporal order was for conveying and storing information in neuronal circuits, and experimenters saw how the synaptic connections of the brain
should be acutely sensitive to timing. Thus the field of  STDP was born, via the key studies of  Markram  and Gerstner~\cite{markram1995action,gerstner1996neuronal}.
With STDP, a neuron embedded in a neuronal network can `determine' which neighbouring neurons are worth connecting with, by potentiating those inputs that predict its own spiking
activity, and effectively ignoring the rest~\cite{markram2012spike}.
The net result is that the sample neuron can integrate inputs with predictive power and
transform this into a meaningful predictive output, even though
the meaning itself is not strictly known by the neuron.

An early example of using such models in associative memory
can be found in~\cite{gerstner1992associative}. This introduces `spiking' neurons in a
Hopfield~\cite{hopfield1982,hopfield1984} network: by the term spiking, three main features are implied, which are: a) a neuron fires when a given {\it threshold} is reached; b) it then undergoes a period of rest,
 which is referred to as `refractoriness'; c) noise may be added to the firing rates. The synapses connecting the neurons follow a Hebbian learning rule
(with no explicit competition) whereby incoming patterns are learnt, and their retrieval analysed along the lines of Section~\ref{hopf} as a function of various parameters.

 While models of neurons themselves are the 
subject of considerable discussion~\cite{gerstner2009good}, these early models have been greatly refined in recent 
times, and are usefully summarised in~\cite{Sjoestroem2010}. However, as pointed out in~\cite{markram2012spike}, these theories are
limited by the types of plasticity invoked in the models concerned. Indeed, in~\cite{zenke2014diverse}, it is tacitly acknowledged that without appropriate compensatory mechanisms 
(referred to there as being `non-Hebbian'),
Hebbian learning alone is not able to account for the reliable storage and recall of memories; the necessary mechanisms invoked in~\cite{zenke2014diverse} involve,
in addition to the Hebbian LTP/LTD, the (implicitly competitive) mechanism of
heterosynaptic up and down regulation of synapses, as well as transmitter-induced plasticity and consolidation.
This indeed reinforces the perceived need for some form of competition, as well as  a somewhat more parsimonious form
of modelling where possible. 

Before concluding, we also mention that 
most STDP models can be averaged and reduced to rate-based
models with certain assumptions: if all nodes interact with each other, they can be reduced to correlation-based models~\cite{kempter} whereas if nearest-neighbour interactions
exist, the models that result are similar to the BCM model~\cite{izhik}.
However, the fast dynamics of neurons, on which the STDP models are based, continue to attract a lot of research interest.
Typically, models of integrate-and-fire neurons on networks have been extensively studied, and their different dynamical regimes
explored~\cite{brunel}. In~\cite{dubreuil2016storing}, the memory performance of a class
of modular attractor neural networks has been examined, where modules are potentially fully-connected networks connected to each other via diluted long-range connections. 
Interest in this fast dynamical regime has also been fuelled by
the discovery of neuronal avalanches in the brain~\cite{beggs}, which was followed by several dynamical models of neural networks~\cite{born,roxin}, where the statistics of avalanches were 
investigated~\cite{lucilla1,lucilla2,geisel1,geisel2,geisel3,taylor} and reviewed in~\cite{rabi}. In fact, the field of spiking neurons is now so well-established that it is the
 subject of textbooks -- of which an excellent
example is the one by two of the most important workers in the field, Gerstner and Kistler~\cite{book3}.

\section{State-based models}\label{state}

An alternative to considering unbounded and continuous synaptic weights -- as is done in Sections~\ref{hopf} and~\ref{phen} -- is to consider {\it discrete} synapses, with a limited
number of synaptic states, whose weights are bounded. This has experimental support~\cite{vanrossum,o2005graded}, and also has the advantage that binary synapses, say, may be more robust to noise than continuous synapses~\cite{crick1984neurobiology}. An essential property of these models as well as real neural networks is that their capacity is finite. 
Such bounded synapses have the palimpsest property, i.e., new memories are stored at the cost of old ones being overwritten~\cite{parisi}. This is  in marked contrast
to the case of unbounded synapses 
where the overall quality of both old and new memories degenerate as new information is processed. For bounded synapses, therefore,
forgetting is an important aspect of continued learning~\cite{nadal,parisi,amit1992constraints,amit1994learning,amit1997dynamics,landf4,vanrossum}.
This situation --  that of {\it discrete, bounded synapses with an explicit
forgetting mechanism} --  is what we will focus on in the rest of this review.

Van Rossum and coworkers~\cite{barrett,barrett2008optimal} have done a body of work on such state-based models; they have shown in particular that there is not an overwhelming
reduction in the storage capacity of discrete synapses as compared to continuous ones. In their work,
each synapse is described with a state-diagram
and each state has an associated synaptic weight. The simplest case of binary synapses (`synaptic switches'),
has been extensively used in earlier mathematical models~\cite{fusi2006eluding,fusi2007limits,p2,barrett}.
Interactions between synapses are incorporated in the state diagrams. Typically, Markov descriptions are used,
and the eigenvalues of the Markov  transition matrices give the decay times of the synaptic weights.

The above mechanism of synaptic plasticity has, however, been shown to be rather inefficient when synapses
change permanently~\cite{amfuss}. Pure plasticity indeed does not provide a mechanism
for protecting some memories while leaving room for other,
newer, memories to come in,
hence leading to the need for the mechanism of metaplasticity~\cite{amit1992constraints,amit1994learning,amit1997dynamics}\footnote{An older use of the term `metaplasticity' relates to changes in synapses that are not expressed as changes in synaptic efficacy, but rather alter their responses to subsequent stimuli, an example of this being the sliding threshold of plasticity described in the BCM model~\cite{bcm}.}. In order to improve performance,
Fusi~\etal~\cite{fusi} proposed a cascade model of a synapse with many hidden states,
which they claimed was able to store long-term memories more efficiently,
with a decay that was {\it power-law} rather than exponential in time.The pathbreaking idea behind the work of~\cite{fusi} was that the introduction
of `hidden states' for a synapse would enable the delinking of
memory lifetimes from instantaneous signal response: while maintaining quick learning,
it would also enable slow forgetting.
In the original cascade model of~\cite{fusi}, this was implemented
by the storage of memories at different `levels': the relaxation times for
the memories increased as a function of depth.
It was assumed that
short-term memories, stored at the uppermost levels, would decay as a
consequence of their replacement by other short-term memories (`noise').
On the other hand, longer-lasting memories remained largely immune to such
noise as they were stored at the deeper levels, which were accessible only rarely.
This hierarchy of timescales models
the phenomenon of metaplasticity~\cite{meta1,meta2}, and will be discussed in detail below.

\subsection{Fusi's cascade model: A quantitative formulation}\label{casc}

Fusi's model~\cite{fusi} 
of a metaplastic binary synapse with infinitely many hidden states was formulated quantitatively and investigated in~\cite{jstat}. Each state is here
labelled by its depth $n=0,1,\dots$, 
At every discrete time step $t$, the synapse is subjected either to
an LTP signal (encoded as $\eps(t)=+1$) or to an LTD signal (encoded as $\eps(t)=-1$),
where $\eps(t)=\pm1$ is the instantaneous value
of the input signal at time $t$.

\begin{figure}[!ht]
  \begin{center}
    \includegraphics[angle=+00,width=.25\linewidth]{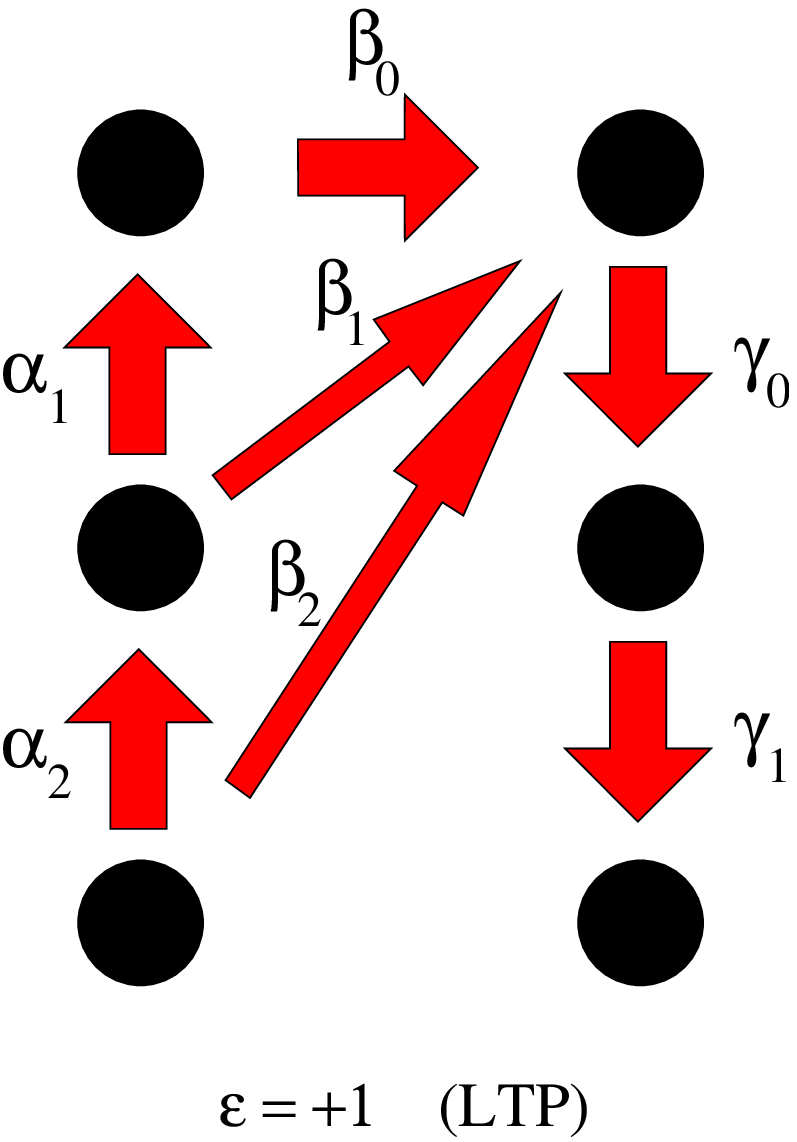}
    {\hskip 25pt}
    \includegraphics[angle=+00,width=.25\linewidth]{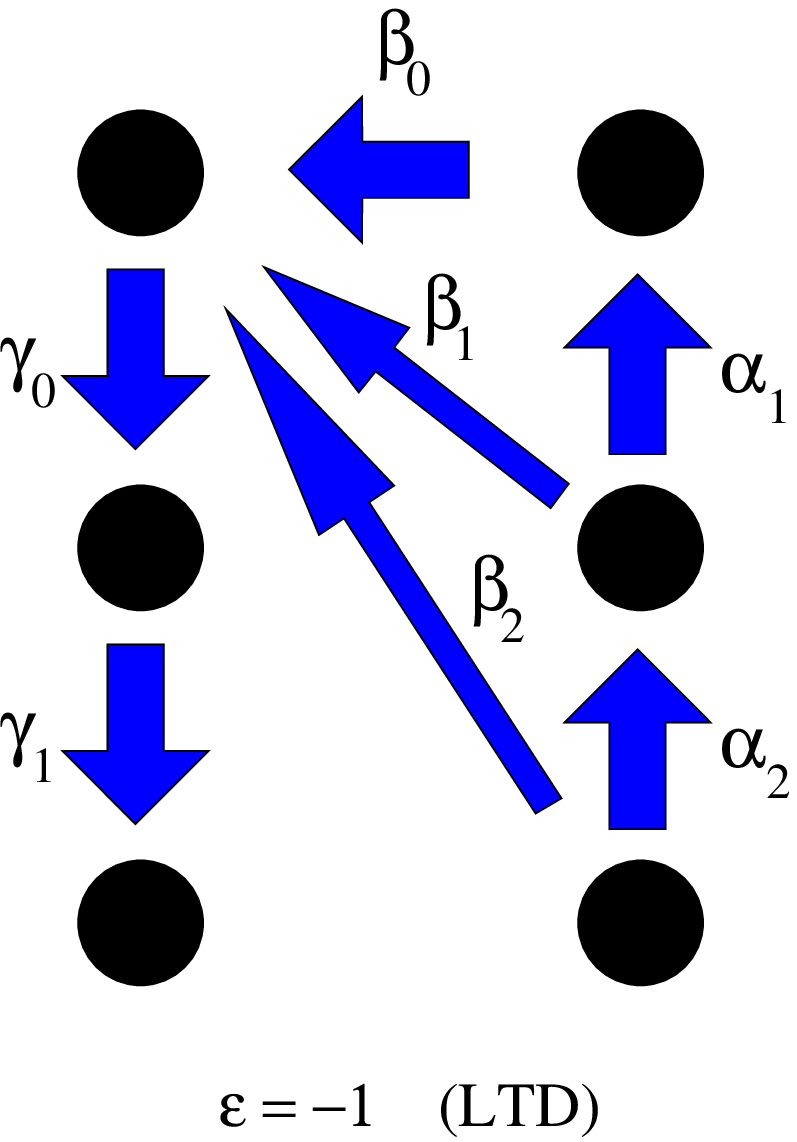}
    \caption{\label{I}
      Schematic representation of Model~I.
      Arrows denote possible transitions in the presence of an LTP signal
      ($\eps=+1$, left panel)
      and of an LTD signal
      ($\eps=-1$, right panel).
      Corresponding transition probabilities are indicated.
      In each panel, the left (resp.~right) column corresponds to the $-$ (resp.~$+$) state.
      The model studied in this work is actually infinitely deep
      (after Ref.~\cite{jstat}).}
  \end{center}
\end{figure}

The model, portrayed in Figure~\ref{I}, is defined as follows:
The application of an LTP signal can have three effects~\cite{jstat}:
\begin{itemize}
\item
  If the synapse is in its $-$ state at depth $n$,
  it may climb one level $(n\to n-1)$ with probability~$\a_n$.
  (This move was absent in the original model.)

\item
  If it is in its $-$ state at depth $n$,
  it may alternatively hop to the {\it uppermost} $+$ state with probability~$\b_n$.

\item
  If it is already in its $+$ state at depth $n$,
  it may fall one level $(n\to n+1)$ with probability~$\g_n$.
\end{itemize}

Long-term memories will be stored in the deepest levels of the synapse,
because of the persistent application of unimodal signals.
The effect of noise on such a long-term memory here is
to replace a long-term memory by a short-term
memory of the opposite kind.
If, for example, the signal is composed of all $+++++++$, an isolated
$-$ event could be seen to represent the effect of noise.
In this case, the
Fusi model~\cite{fusi} predicts that the signal is thrown from a deep positive level of
the synapse to the uppermost level of the negative pole.
Seen differently,
this mechanism converts a long-term memory of one kind to a short-term
memory of the opposite kind.

Along the lines of~\cite{fusi, jstat},
the transition probabilities of this model
are assumed to decay exponentially with level depth $n$:
\beq
\a_n=\a\e^{-(n-1)\mud},\qquad
\b_n=\b\e^{-n\mud},\qquad
\g_n=\g\e^{-n\mud}.
\label{rates}
\eeq
The corresponding characteristic length,
\beq
\xid=\frac{1}{\mud},
\eeq
is one of the key ingredients of the model,
which measures the number of fast levels at the top of the synapse.
It will be referred to as the {\it dynamical length} of the problem.
The choice made in~\cite{fusi} corresponds to $\e^{-\mud}=\half$, i.e.,
$\mud=\ln 2$.
A different characteristic length, the static length $\xis$,
is given by
\beq
\xis=\frac{1}{\mus}.
\eeq
This is referred to as the {\it static length} of the problem, and
gives a measure of the effective number of occupied levels in the default state~\cite{jstat}.
The regime of most interest is where~$\xis$ is moderately large,
so that the default state extends over several levels.
The mean level depth
\beq
\mean{n}^{\text{st}}=\frac{1}{\e^\mus-1}=\xis-\half+\cdots
\label{meanstat}
\eeq
is then essentially given by the static length.

The {\it level-resolved output signal} of level $n$ at time $t$:
\beq
D_n(t)=Q_n(t)-P_n(t)
\label{dndef}
\eeq
and the {\it total output signal} at time $t$:
\beq
D(t)=\sum_{n\ge0}D_n(t)
\label{ddef}
\eeq
can be expressed in terms of the probabilities $P_n(t)$ (or $Q_n(t)$)
for the synapse to be in the $-$ state (or the $+$ state)
at level $n=0,1,\dots$ at time $t=0,1,\dots$

%When a single potentiating pulse signal is applied at time $t=1$ (that is, $\eps(1)=+1$)
%to the synapse in its default state.
%the synapse will get polarized in response, and thus `learn' the signal.
%Later on, under the influence of a random input signal for times $t\ge2$,
%it will `forget' the PP signal, and return to its default state.
%Figure~\ref{ltpred} shows plots of the reduced output signal $D(t)/D(1)$
%against time $t$ for several values of the control parameter $\b$.
%All subsequent figures refer to the parameter values $\b=0.2$, $\g=0.5$,
%and $\xis=\xid=5$ (see last section).
%From here on, we will refer to times where the synapse is subjected to a
%significant signal ($\eps(t)=\pm1$) as {\it learning phases},
%and to times where the synapse is subjected to random input
%($\eps(t)=0$) as {\it forgetting phases}.

%If the synapse were finite rather than infinite, and consisted of $N$ levels,
%the power-law decay (\ref{dtheta}) would be exponentially cut off at a time
%\beq
%\tau_N\sim\exp(N/\xid)
%\eeq
%which grows exponentially fast with the ratio
%of the number $N$ of levels to the dynamical length $\xid$.

%\begin{figure}[!ht]
%  \begin{center}
%    \includegraphics[angle=-90,width=.3\linewidth]{ltpred.eps}
%  \end{center}
%  \caption{
%    Plot of the reduced output signal $D(t)/D(1)$ after a single PP input signal,
%    against time $t$, for several $\b$.
%    After~\cite{jstat}.}
%  \label{ltpred}
%\end{figure}

We now describe the effect of an LTP signal, i.e., a sustained input of potentiating pulses
lasting for $T$ consecutive time steps ($\eps(t)=+1$ for $1\le t\le T$)
on the model synapse.
The synapse, assumed to be initially in its default state~\cite{jstat}
 will get almost totally polarized in response to the persistent signal.

\begin{figure}[!ht]
  \begin{center}
    \includegraphics[angle=-90,width=.45\linewidth]{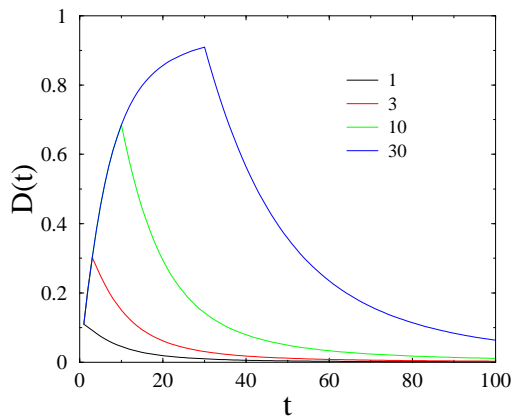}
  \end{center}
  \caption{
    Plot of the output signal $D(t)$
    against time $t$, for several durations $T$ of the LTP signal for parameter values $\b=0.2$, $\g=0.5$,
   and $\xis=\xid=5$ 
    (after Ref.~\cite{jstat}).}
  \label{sted}
\end{figure}

This saturation phenomenon is illustrated in Figure~\ref{sted},
which shows the output signal $D(t)$
for several durations $T$ of the LTP signal.
The synapse slowly builds up a long-term memory in the presence
of a long enough LTP signal, as the memorized signal moves to deeper
and deeper levels.
At the end of the learning phase ($t=T$),
the polarisation profile will have the form of a sharply peaked traveling wave,
around a typical depth which grows according to the logarithmic law~\cite{jstat}
\beq
n(T)\approx\xid\ln\g T.
\label{nT}
\eeq

After the signal is switched off, the total output signal  decays.
The late stages of the forgetting process are characterized by a universal power-law decay of the output signal:
\beq
D(t)\sim t^{-\theta}.
\label{dtheta}
\eeq
This is known as {\it power-law forgetting}~\cite{wixted1991form,wixted1997genuine,plf3}.
The forgetting exponent
\beq
\theta=1+\frac{\xid}{\xis}
\label{theta}
\eeq
is always larger than unity and depends on the ratio of the dynamical and static lengths
$\xid$ and $\xis$. As Equation~(\ref{dtheta}) shows, it has no dependence on the
the duration of the learning phase, in keeping with the requirements of universality.

\subsection{Comparison of cascade model with experiment}

The cascade model and its variants have frequently been criticised for being somewhat abstract; one response has been
to come up with ever-more sophisticated models for synaptic consolidation which incorporate the multiple timescales inherent in the cascade model.
A three-layered model of synaptic
consolidation has been proposed that accounts for data across a large range of experimental conditions~\cite{ziegler2015synaptic}; while it
 has a daunting number of parameters -- 17 --, it is able to incorporate the retention of long-term memories. 
Fusi's own recent
extension of the cascade model is also rather intricate: memories are stored and retained through complicated coupled processes operating on multiple timescales. This is achieved by combining multiple dynamical processes that initially
store memories in fast variables, and then progressively transfer them to slower variables. It has the advantage of getting a larger memory capacity, while the corresponding disadvantage is that it is even more abstract than his
earlier model, so that involved biological processes have to be explained via systems of communicating vessels~\cite{benna2016computational}. 

We choose here instead to highlight a link  with an experiment~\cite{luk} whose findings are explained by the complex synaptic architectures
of Fusi's original model~\cite{fusi}, to combat the proposition that the cascade model is `too abstract' to be useful. In particular the experiment
involves a single synapse connecting two cells, so that the Fusi model of a single synapse is appropriate.
Specifically,
in a system comprising an
excitatory 
synapse between {\it Lymnaea} pre- and postsynaptic neurons (visceral dorsal 4 (VD4) and left pedal dorsal 1 (LPeD1-
Excitatory)), a novel form of short-term potentiation
was found, which was use-, but not
time-dependent~\cite{luk}.
Following a tetanic stimulation ($\sim$ 10 Hz) in the presynaptic
neuron with a minimum of seven action potentials,
the synapse became potentiated whereby a subsequent action
potential triggered in the presynaptic neuron resulted in an enhanced
postsynaptic potential.
Further, if an inducing tetanic stimulation was activated,
but a subsequent action potential was {\it not} triggered,
the synapse was shown to remain potentiated for as long as 5 hours.
However, once this action potential was triggered,
the authors found that the synaptic strength rapidly returned to baseline levels.
It was also shown that this form of synaptic plasticity relied on the
presynaptic neuron, and required pre- (but not post-) synaptic
Ca$^{2+}$/calmodulin dependent kinase II (CaMKII) activity.
Hence, this form of potentiation shares induction and de-potentiation
characteristics similar to other forms of short-term potentiation,
but exhibits a time-frame analogous to that of long-term potentiation.

In~\cite{plosone}, this experiment was interpreted via a variant of the cascade model described
above, as follows:
after a process of tetanic stimulation,
the initial action potentials, interpreted as a non-random signal,
cumulatively built up a long-term
memory of the signal in the deepest synaptic levels.
The synapse dynamics were then frozen so that further discharge was prevented.
When a further action potential was applied, the synaptic dynamics restarted (`use'-dependence):
the release of the accumulated memory from the deepest levels of the synapse
constituted the observed enhancement of the output signal described in~\cite{luk}.
While this enhancement is plausibly accounted for by
the model of metaplastic synapses~\cite{jstat}, the explanation of the
freezing of the synaptic dynamics and its subsequent use-dependence needed the introduction of a stochastic and bistable biological switch to model the role of kinase
(CaMKII) in the actual experiment~\cite{luk}.

Specifically, the synapse (Figure~\ref{I}), assumed to be initially in its default state,
is subjected to a sustained LTP signal of duration $T_1$
(i.e., the application of $T_1$ action potentials),
and to a single action potential at a much later time ($T_2\gg T_1$).
It is subjected to a random input
at all the other instants of time
($\eps(t)=+1$ for $1\le t\le T_1$ and for $t=T_2$, else $\eps(t)=0$).
In the regime where the number of action potentials $T_1$ of the initial signal
is larger than some characteristic time $T_0$ of the switch,
the freezing probability of the switch at the end of the LTP period is very high,
i.e., very close to unity.
During this learning phase,
the output signal $D(t)$ grows progressively from $D(0)=0$
to a large value $D(T_1)$.
The high value of the freezing probability at the end of this phase
typically freezes the synaptic dynamics,
ensuring that this enhanced output signal is not discharged.
When the next action potential is applied at time $T_2$, the switch is turned off,
and the synapse then relaxes via
the full discharge of the stored, enhanced output signal.

\begin{figure}[!ht]
  \begin{center}
    \includegraphics[angle=-90,width=.48\linewidth]{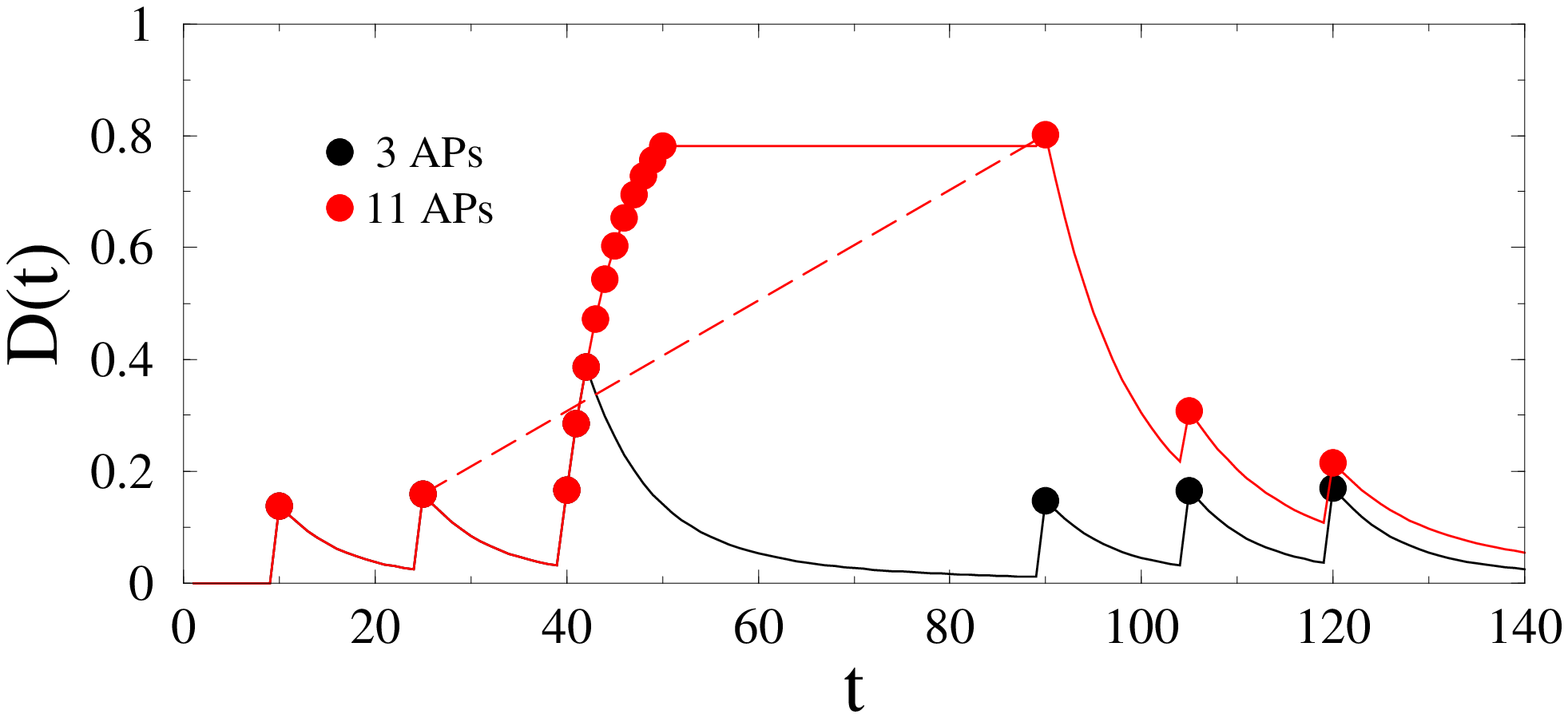}{\hskip 11pt}

    {\hskip 12pt}\includegraphics[angle=0,width=.5\linewidth]{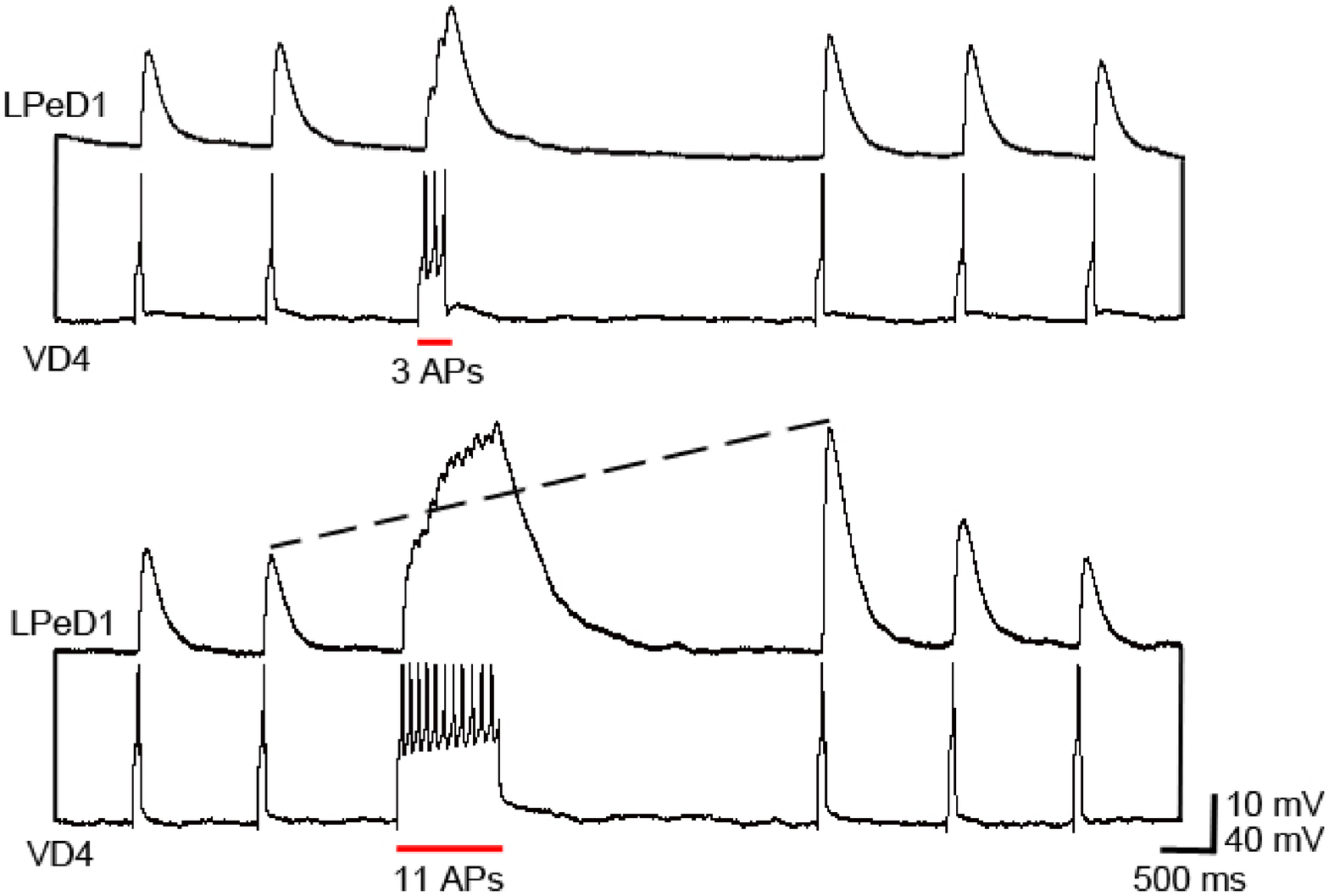}
  \end{center}
  \caption{
    An integrative figure showing the predictive model (upper panel)
    and sharp-electrode electrophysiology recordings of a VD4/LPeD1 synaptic pair
    (two lower panels).
    While three action potentials triggered during tetanic stimulation
    are insufficient to result in potentiation of a subsequent
    excitatory postsynaptic potential (EPSP) in the LPeD1 neuron,
    eleven action potentials elicited during tetanic stimulation result
    in a potentiated response, as predicted by the model
    (after Ref.~\cite{plosone}).}
  \label{FigureX}
\end{figure}

Figure~\ref{FigureX} shows a quantitative comparison between
the theoretical predictions of~\cite{jstat} (upper panel)
with sharp-electrode electrophysiology recordings
of a VD4/LPeD1 synaptic pair (two lower panels)~\cite{plosone}.
The black theoretical curve corresponds to
3 APs triggered during tetanic stimulation,
which are insufficient to result in potentiation of a subsequent
excitatory postsynaptic potential (EPSP) in the LPeD1 neuron
($T_1=3\ll T_0$, so that the switch remains off).
The red theoretical curve corresponds to 11 APs,
resulting in a potentiated subsequent response
($T_1=11\gg T_0$, so the switch is turned on and the synapse is frozen).
The model biological switch used to model the action of kinase in~\cite{plosone} displays an essential bistability
so that the phenomenon described above is observed more or less frequently
depending on the difference between
 the duration $T_1$ of the initial LTP signal
and the characteristic time $T_0$ of the switch.

%\begin{figure}[!ht]
%  \begin{center}
%    \includegraphics[angle=-90,width=.3\linewidth]{switch1.eps}
%    {\hskip 10pt}
%    \includegraphics[angle=-90,width=.3\linewidth]{switch2.eps}
%  \end{center}
%  \caption{
%    Plot of the two possible kinds of output signals $D(t)$
%    generated by the protocol described in the text, against time $t$
%    ($T_0=5$, $T_2=50$).
%    Symbol sizes are proportional to the probabilities of each kind of behavior,
%    i.e., $\Pi(T_1)$ for the frozen one and $1-\Pi(T_1)$ for the unfrozen one.
%    Left: $T_1=7$ is larger than $T_0=5$, so that $\Pi(T_1)=0.946$ is very high.
%    Right: $T_1=T_0=5$, so that $\Pi(T_1)=\half$.}
%  \label{switch}
%\end{figure}

Thus, despite its seeming abstraction, the basic ideas of Fusi's cascade model can indeed be related
to real experimental data; in fact, such complex synaptic architectures provide fertile ground for the inclusion
of multiple timescales which are essential to the modelling of long-term memory.

\section{Synaptic dynamics: the need for competition}\label{comp}

In the models of the preceding section, while synapses have been central to the acquisition and recall of long-term memory,
there has been no mention of their embedding networks, in particular to do with the neurons that synapses connect. In this section
we return to the concepts of Section~\ref{hopf}, and to the explicit mechanisms of synaptic strengthening and weakening that result
from neuronal firing within a network. 
We have already discussed in Section~\ref{phen} several phenomenological models of synaptic plasticity, where the need for competitive
dynamics has been made clear. In the following, we elaborate on several ways in which these have been implemented in the neuroscience literature.

In the following, we follow the lines of argument of
Van Ooyen's excellent review article on synaptic competition~\cite{ooyen2001competition},
 where a distinction is first made between {\it independent} and {\it interdependent competition}.
In {\it interdependent} competition, victors emerge as a result of interactions between  participants, such as in a sporting event.
Interdependent competition is frequently considered, for example, in
population biology; here,
two species are said to compete if they try to limit
the growth of each others' population.
In  {\it independent} competition, on the other hand, the participants do not interact,
but are rather chosen on the basis of some sort of contest.
This kind of competition is reminiscent of
{\it competitive learning}  which was introduced by Kohonen~\cite{kohonen1982self}, 
and which will form the basis of the rest of this article.

In neural network models based on competitive learning, 
only synapses connected to the neurons most responsive to stimuli have
their strengths changed. What is implicit here is that these stimuli come from presynaptic neurons
so that their {\it correlated} transmission to postsynaptic neurons causes the corresponding synapses to be strengthened~\cite{swindale1996development}.
Such synaptic competition~\cite{avy}  often arises through Hebbian learning
so that when the synaptic strength of one input grows, the strength of the others shrinks. Whereas many models
phenomenologically enforce competition by requiring the total strength of all synapses onto a postsynaptic
cell to remain constant~\cite{miller}, others implement biochemical processes and modified Hebbian learning rules.

To see how competition between input connections can be enforced, consider $n$ inputs,
with synaptic strengths $J_i (t) (i = 1, {\dots}, n)$, impinging on a given postsynaptic cell at time~$t$.
Simple Hebbian rules for the change 
$\mathup{\Delta} J_i(t)$ 
in synaptic strength in a time interval 
$\Delta t$ state that
the synaptic strength should grow in proportion to the product of the postsynaptic activity level
$y(t)$ and the activity level $x_i (t)$ of the $i$th input. Thus
\begin{equation}
  \mathup{\Delta} J_i(t)\propto y(t)x_i(t)\mathup{\Delta} t\:.
\label{synnorm}
\end{equation}
If two inputs activate a common target, one needs competition to make one of the synaptic strengths grow at the expense of the other.
A common method to achieve this is to constrain the total synaptic strength
 via {\it synaptic normalisation} --  this is the constraint that
\beq
\sum_i^n J_i^p(t)=K,
\label{synnormk}
\eeq
with $K$  constant and the integer $p$  usually taken to be $1$ or $2$. Specifically, $p = 1$ conserves the total synaptic strength, whereas $p = 2$ conserves the length of the weight vector. At each
time interval 
$\Delta t$, following a phase of Hebbian learning, in which
$J_i(t+\mathup{\Delta})=J_i(t)+\mathup{\Delta}J_i(t)$,
the new synaptic strengths are forced to satisfy the normalization constraint of Equation~(\ref{synnormk}). 
Typically this can be enforced by one of two processes: {\it multiplicative} or {\it subtractive} normalisation. These ensure that synaptic strengths do not grow without bounds.

In {\it subtractive} normalization~\cite{miller1989ocular,miller1990development},
the same amount is subtracted from each weight to enforce the
constraint. In
{\it multiplicative} normalization~\cite{malsburg1973self,von1976mechanism,willshaw1976patterned,willshaw1979marker} on the other hand, each synaptic weight $J_i (t + \Delta 
t)$ is scaled in proportion to its size. A two-layer
model is there proposed, where the stimuli in neurons of  the input layer are sent to an  output layer 
of neurons.
If the neuronal inputs are above some specified threshold, then the responses in the output layer are calculated, taking into account
the pattern of synaptic connections; weights are updated by a Hebbian rule after this neuronal activity stabilises.
The final outcome of development may of course differ depending on whether multiplicative
or subtractive normalization is used~\cite{miller1994role, simpson2009theoretical}.

Kohonen~\cite{kohonen1982self}
proposed
a drastic but effective simplification of the
approach of~\cite{malsburg1973self}. 
In the latter, a few hotspots of activity typically emerged in the output layer
following the iterations of the input activity via the lateral synapses. To obviate the
considerable time taken to ensure the convergence of these iterations,
Kohonen proposed the centering of the activity in the output layer on the so-called `winning' neurons, followed by
 standard Hebbian learning. This important simplification is vital to the statistical physics approaches that will be
presented in Section~\ref{complearn}. Another way of viewing this is to regard it as yet another nonlinear
approach to competitive learning; if the layer of output neurons is assumed to be connected by {\it inhibitory} synapses,
the neuron with the largest initial activity can be said to suppress the activity of all other output neurons. 

The competitive approaches described in the above paragraphs are often described as
{\it hard}, in the sense of being `winner-take-all'. In {\it soft} competitive
learning, all neurons in the output layer are updated by an amount that
takes into account both their feed-forward activation and the activity of
other output neurons. This will also be seen to have equivalences with {\it agent-based learning
models} in the statistical physics approaches of Section~\ref{complearn}.

Another approach for achieving competition 
 is to modify the simple Hebbian learning rule 
of Equation~(\ref{synnorm})
so that both increases in
synaptic strength (LTP) and decreases in synaptic strength (LTD) can take place.
If we assume that the presynaptic activity level $x_i(t)$ 
as well the postsynaptic activity level $y (t)$ 
must be above some thresholds, respectively 
$\theta_x, \theta_y$,
to achieve LTP  (and otherwise yield LTD),
then a suitable synaptic modification rule is~\cite{miller}
\begin{equation}\label{eq:Ooyen2}
  \mathup{\Delta}J_i(t)\propto [y(t)-\theta_y][x_i(t)-\theta_x]\mathup{\Delta}t\:.
\end{equation}

Thus, if both $y(t)$ and $x_i (t)$ are above their respective thresholds, LTP occurs; if one
is below its threshold and the other is above, LTD occurs. 
For this to qualify as proper competition,
 the synaptic strength lost through LTD must roughly equal the strength
gained through LTP. This can only be achieved with appropriate input correlations, which
makes simple LTD a fragile mechanism for achieving competition~\cite{miller}.
Another mechanism which ensures that when some synaptic strengths increase, others must
correspondingly decrease - so that competition occurs - is to make one of the thresholds variable. If
the threshold  $\theta_x^i$
increases sufficiently as the postsynaptic activity $y(t)$ 
or synaptic strength
$J_i(t)$ 
increases, conservation of synaptic strength is achievable~\cite{miller}.
Similarly, if the threshold 
$\theta_y$
increases faster than linearly with the average postsynaptic activity,
then the synaptic strengths will adjust to keep the postsynaptic activity near a limiting value~\cite{bcm}. This, however,  results in temporal competition between input patterns, rather
than spatial competition between different sets of synapses.

So far, causal links between seemingly correlated firings of neurons have been assumed. As before, 
spike-time dependent plasticity (cf. Section~\ref{stdp}) makes this explicit via
its emphasis on the
the relative timing of
pre- and post-synaptic activity. 
In the approach of~\cite{song},
presynaptic activity that precedes postsynaptic spikes strengthens a synapse,
whereas presynaptic activity that follows postsynaptic spikes, weakens it. 
As a consequence of the intrinsic nonlinearity of the spike generation
mechanisms,
and with the imposition of hard limits on synaptic strengths,
STDP has the effect of keeping the total synaptic input to the neuron
roughly constant, independent of the presynaptic firing rates.
This approach, of rewarding truly correlated neuronal activity while penalising its absence,
has been taken into account in the models of synaptic dynamics presented in Section~\ref{modelsyndyn}.

\section{Statistical physics models of competing synapses}\label{statsyn}

The emergence of new areas in physics has strongly contributed to the development of analytical tools; this is particularly true for the
field of complex systems. A particular area which is of relevance in the context of this review is that of {\it agent-based modelling}; here, local interactions among agents may give rise to
 emergent phenomena on a macroscopic scale~\cite{santo}. In these models, agents on the sites of appropriately defined lattices interact with each other; their collective behaviour is then analysed
in terms of global outcomes. A typical example arises in, say, the context of financial markets; trading rules between different agents at an individual level can result in specific sets of traders, or
their representative strategies, winning over their competitors. This makes for interesting analogies with {\it competitive learning}; approaches based on this have therefore successfully been
 used to investigate a wide variety of topics, ranging
from the diffusion of innovations~\cite{uspr, kchat} through gap junction connectivity in the pancreas~\cite{pranme} to the dynamics of {\it competing} synapses~\cite{gmam1, gmam2, luck2014slow}. It
is the latter which will concern us here, but in the interests of completeness, we first briefly review an agent-based model of competitive learning in the following~\cite{uspr}.

\subsection{An agent-based model of competitive learning}\label{complearn}

The underlying idea~\cite{uspr} is that the strategy of a
given agent is to a large extent determined by what the other agents
are doing, through considerations of the relative payoffs obtainable in each case.
 Agents are located at the sites
of a regular lattice, and can be associated with one of
two types of strategies.
Every
agent revises its choice of type at regular intervals, and in
this it is guided by two rules: a majority rule, reflecting
the tendency of  agents to align with their local
neighborhood, followed by an adaptive performance-based
rule, via which the agent chooses the type that
is more successful locally.

Assuming that the agents sit at the nodes of a \textit{d}-dimensional regular lattice with coordination number $z = 2d$, the efficiency of an agent at site \textit{i} is represented by an Ising spin variable:
\begin{equation}\eta_{i}(t) = \begin{cases}
    +1 & \mbox{ if \textit{i} is $+$ at time \textit{t},} \\
    -  1 & \mbox{ if \textit{i} is $-$ at time \textit{t}.}
  \end{cases}
\label{gameth1}
\end{equation}
The evolution dynamics of the lattice is
governed by two rules. The first is a \textit{majority} rule, which consists of the alignment of an agent with the local field (created by its nearest neighbours) acting upon it, according to:
\begin{equation}
  \eta_{i}(t+\tau_{1})=\left\lbrace\begin{array}{cccc}+1 &  & \mbox{if}\ & h_{i}(t)> 0,\\\pm 1  & \makebox[1.5cm]{w.p.}\frac{1}{2} {\hskip 6pt} & \mbox{if}\ & h_{i}(t)=0,\\-1 &    &  \mbox{if}\ & h_{i}(t)<0. \end{array} \right.
\end{equation}
Here, the local field
\begin{equation}
  h_\textit{i}(t) = \sum_{\textit{j(i)}}{\eta_\textit{j}(t)}
\label{gameth2}
\end{equation}
is the sum of the efficiencies of the $\textit{z}$ neighbouring agents $\textit{j}$ of site $\textit{i}$ and $ \tau_{1}$ is the associated time step.
Next, a \textit{performance} rule is applied. This starts with the assignment of an outcome $\sigma_{i}$ (another Ising-like variable, with values of $\pm 1$ corresponding to success and failure respectively) to each  site $\textit{i}$, according to the following rules:
\begin{eqnarray}
  \mbox{if}\  \eta_{i}(t)= +1,\nonumber\\
  \mbox{then } & \sigma_{i}(t+\tau_{2})=\left\lbrace\begin{array}{lll}+1 & \mbox{w.p.} & p_{+},\\-1\  & \mbox{w.p.}\  & 1- p_{+},\end{array}\right.\nonumber\\
  \mbox{if}\  \eta_{i}(t)= -1,\nonumber\\
  \mbox{then } & \sigma_{i}(t+\tau_{2})=\left\lbrace\begin{array}{lll}+1 & \mbox{w.p.} & p_{-},\\-1\  & \mbox{w.p.}\  & 1- p_{-},\end{array}\right.
\end{eqnarray}
where $\tau_{2}$ is the associated time step and $p_\pm$ are the probabilities of having a successful outcome
for the corresponding strategy. With $N_{i}^{+}$ and $N_{i}^{-}$  denoting  the total number of neighbours of a site \textit{i} who have adopted strategies $+$  and $-$  respectively, and  $I_{i}^{+}$ ($I_{i}^{-}$) denoting the number of successful outcomes within the set  $N_{i}^{+}$ ($N_{i}^{-}$), the dynamical rules for site \textit{i} are:
\begin{eqnarray}
  \mbox{if}\ \eta_{i}(t)=  +1  & {\hskip -60pt} \mbox{ and} {\hskip 15pt} \frac{I_{i}^{+}(t)}{N_{i}^{+}(t)}<\frac{I_{i}^{-}(t)}{N_{i}^{-}(t)},\nonumber\\
  \mbox{then } & \eta_{i}(t+\tau_{3})=\left\lbrace\begin{array}{lll}-1 & \mbox{w.p.} & \varepsilon_{+}\\+1 & \mbox{w.p.} & 1- \varepsilon_{+},\end{array}\right.\nonumber\\
  \mbox{if}\ \eta_{i}(t)=  -1  & {\hskip -60pt} \mbox{ and} {\hskip 15pt} \frac{I_{i}^{-}(t)}{N_{i}^{-}(t)}<\frac{I_{i}^{+}(t)}{N_{i}^{+}(t)},\nonumber\\
  \mbox{then } & \eta_{i}(t+\tau_{3})=\left\lbrace\begin{array}{lll} +1 & \mbox{w.p.} & \varepsilon_{-}\\ -1 & \mbox{w.p.} & 1- \varepsilon_{-}.\end{array}\right.
\label{gamethn}
\end{eqnarray}
Here, the ratios $\frac{I_{i}(t)}{N_{i}(t)}$ are nothing but the average payoff assigned by an agent to each of the two strategies in its neighbourhood at time $t$ (assuming that success yields a payoff of unity and failure, zero). Also, $\tau_{3}$ is the associated time step and the parameters $\varepsilon_{\pm}$ are indicators of the memory associated with each strategy. In their full generality, $\varepsilon$ and~$p$ are independent variables: the choice of a particular strategy can be associated with either a short or a long memory.

Setting the timescales 
\begin{equation}
  \tau_{2} \rightarrow 0,\quad \tau_{1} = \tau_{3} = 1, 
\end{equation}
the above steps of the performance rule are recast  as effective dynamical rules involving the efficiencies $\eta_{\it{i}}(t)$ and the associated local fields alone:
\begin{eqnarray}
  \mbox{if}\  \eta_{i}(t)= +1,\nonumber\\
  \mbox{then} &  \eta_{i}(t+1) &= \left\lbrace\begin{array}{lll}+1\  & \mbox{w.p.} &\mbox{$w_+$[$h_{i}(t)$]}\\-1\  & \mbox{w.p.}\  &\mbox{1-$w_+$[$h_{i}(t)$]},\end{array}\right.\nonumber\\
  \mbox{if}\  \eta_{i}(t)= -1,\nonumber\\
  \mbox{then} &  \eta_{i}(t+1) &= \left\lbrace\begin{array}{lll}+1\  & \mbox{w.p.} &\mbox{$w_-$[$h_{i}(t)$]}\\-1\  & \mbox{w.p.}\  &\mbox{1-$w_-$[$h_{i}(t)$]}.\end{array}\right.
\end{eqnarray}
The effective transition probabilities $w_{\pm}(h)$ are evaluated by enumerating the $2^{z}$ possible realizations of the outcomes $\sigma_{j}$ of the sites neighbouring site \textit{i}, and weighting them appropriately. 
The specific transition probabilities computed will depend on the embedding lattice of network chosen~\cite{uspr}.

%For a 2D square lattice, the possible local field values at the interfacial sites are 0 and $\pm$2. The corresponding transition probabilities for these field values are~\cite{uspr}:
%\begin{eqnarray}
%  w_{+}(+2) &=& 1 - \varepsilon_{+}p_{-}(1-p_{+}^3),\nonumber\\
%  w_{-}(+2) &=& \varepsilon_{-}(1 - p_{-})[1 -(1-p_{+})^3],\nonumber\\
%  w_{+}(0) &=& 1 - \varepsilon_{+}p_{-}(1 - p_{+})(2 - p_{-} - 2p_{+} + 3p_{-}p_{+}),\nonumber\\
%  w_{-}(0) &=& \varepsilon_{-}p_{+}(1 - p_{-})(2 - p_{+} - 2p_{-} + 3p_{-}p_{+}),\nonumber\\
%  w_{+}(-2) &=& 1 - \varepsilon_{+}(1 - p_{+})[1 -(1-p_{-})^3],\nonumber\\
%  w_{-}(-2) &=& \varepsilon_{-}p_{+}(1-p_{-}^3)\nonumber.\\
%  \label{teqns}
%\end{eqnarray}

The above rules are appropriate for cases where the majority rule is clearly definable, i.e., where there is a mix of agent types. The situation is less clear
when there are large areas of a single species, since then, at least with a sequential update, there is a tendency for any exceptions to revert to the majority type,
whatever their performance. The way around this in~\cite{uspr} was to formulate a so-called `cooperative' model, where, say, a more successful agent surrounded by neighbours
who had failed,
was able to convert all of them to the more successful type, thus stabilising his own success. This {\it hard} rule is like the `winner-takes-all' model of synaptic competition alluded to earlier in this review; 
analogously to that case, there is also a {\it soft} rule, where, while a significant majority of agents were coerced into changing their type, not all were so obliged. In~\cite{uspr} all these models
were explored via ordered sequential updates of the agents, and phase diagrams of their extremely different dynamical behaviour in various regimes were presented. The agents were there also deemed to be memoryless,
i.e., they did not take earlier results into account when they made their choices. These restrictions were progressively removed
in~\cite{bhatme, gmam3}, so that the behaviour of the model with different updates, different levels of memory, as well as different interactions was explored. 

\subsection{A minimal model of synaptic dynamics with emergent long-term memory}\label{modelsyndyn}

The diligent reader will have noted the resemblance between Equations~(\ref{gameth1})
and~(\ref{gamethn}) above,
and some of the equations governing neuronal and synaptic dynamics earlier presented in this paper. Indeed, the detailing of the agent-based model of competitive learning~\cite{uspr}
was to motivate just such a comparison. For example, 
 neuronal firings
 are subject to the kind of local field embodied by Equation~(\ref{gameth2}); the performance in both cases (successful neuronal firings and successful outcomes in the model of~\cite{uspr}) in turn
lead to other dynamical changes, and result in global outcomes. These were precisely the lines of thought that led to the use of such agent-based models of competitive learning in some of the early,
and somewhat simple-minded, models of synaptic dynamics~\cite{gmam1, gmam2}. 

Let us now recall what is needed for a minimal model of memory, via synaptic 
dynamics. 
Both cooperation and competition are needed for a meaningful model of synaptic plasticity~\cite{miller},
with competition acting as a check
on the unstable growth of synaptic weights when cooperation alone is invoked~\cite{turr1,turr2}.  
Since synapses have finite storage capacities, one should also
include
 a representation of the spontaneous relaxation of synapses
when space is created via the spontaneous decay of old memories (cf. the palimpsest effect~\cite{nadal,parisi}).
This is indeed what is done in the model network of synapses and neurons~\cite{luck2014slow} that we will describe in the following.
Like the Fusi~\cite{fusi} model,
it is a model of discrete rather than continuous synapses; unlike it, however, here, there are
explicit mechanisms of synaptic weight change via mechanisms of {\it competing and cooperating synapses} that depend intimately on neuronal firing rates.

The dynamical regime chosen in~\cite{luck2014slow} is that of slow synaptic dynamics,
where neuronal firings are considered stochastic and instantaneous; the synapses `see' only the mean firing
rates of individual neurons, characterising them as active or inactive, on that basis.
As a result of this temporal coarse-graining,
the overall effect of the microscopic noise can be represented
by spontaneous relaxation rates from one type of synaptic strength to the other, so far as the palimpsest mechanism is concerned.
{\it Cooperation} between synapses is incorporated via the usual Hebbian viewpoint, while the most crucial
and original part of the formalism involves synaptic {\it competition} where, along the lines of Kohonen's arguments~\cite{kohonen1982self},
synapses
are converted to the type most responsible for neural activity in their
neighbourhood~\cite{gmam1,gmam2}.

\begin{figure}[!ht]
  \begin{center}
    \includegraphics[angle=-90,width=.4\linewidth]{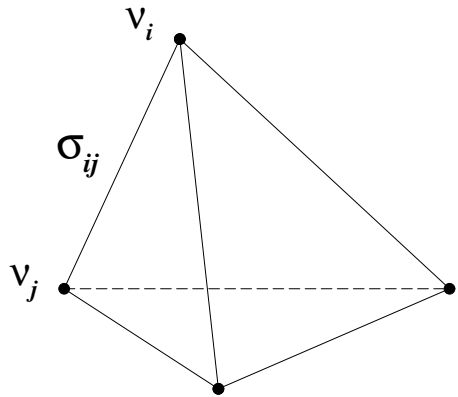}
    \caption{The fully connected network for $N=4$.
      Neurons with activities~$\nu_i=0,\,1$ live on the nodes.
      Synapses with strength types $\s_{ij}=\pm1$ live on the bonds
      (after Ref.~\cite{luck2014slow}).}
    \label{meanfield}
  \end{center}
\end{figure}

The choice of basis is that of a fully connected network,
as depicted in Figure~\ref{meanfield},
so that mean-field theory applies in the thermodynamic limit
of an infinitely large network. 

Neurons live on the nodes (sites) of the network, labelled $i=1,\dots,N$.
The activity state of neuron $i$ at time $t$ is described
by a binary activity variable:
\beq
\nu_i(t)=\begin{cases}
  1&\mbox{if $i$ is {\it active}{\hskip 9pt} at time $t$},\hfill\cr
  0&\mbox{if $i$ is {\it inactive} at time $t$}.\hfill\cr
\end{cases}
\label{nudef}
\eeq
Active neurons are those whose instantaneous firing rate
exceeds some threshold.

Synapses live on the undirected bonds of the network.
The synapse $(ij)$ lives on the bond joining nodes $i$ and $j$.
The strength $J_{ij}$ of synapse $(ij)$ at time $t$ is also described
by a binary variable:
\beq
\s_{ij}(t)=\begin{cases}
  +1&\mbox{if $(ij)$ is {\it strong} at time $t$},\hfill\cr
  -1&\mbox{if $(ij)$ is {\it weak}{\hskip 5pt} at time $t$}.\hfill\cr
\end{cases}
\label{sigdef}
\eeq
Strong synapses are those whose strength $J_{ij}(t)$
exceeds some threshold.

\subsubsection*{Neuronal dynamics}

Neurons have an instantaneous stochastic response to their environment.
The activity of neuron $i$ at time $t$ reads
\beq
\nu_i(t)=\begin{cases}
  1&\mbox{w.p. $F(h_i(t))$},\hfill\cr
  0&\mbox{w.p. $1-F(h_i(t))$},\hfill\cr
\end{cases}
\label{nudyn}
\eeq
where $F(h)$ is an increasing response function
of the input field $h_i(t)$.
The latter is a weighted sum of the instantaneous activities
of all other neurons:
\beq
h_i(t)=\frac{1}{N-1}\sum_{j\ne i}(a+b\s_{ij}(t))\nu_j(t).
\eeq
Strong synapses ($\s_{ij}=1$) enter the sum through a synaptic weight $a+b$,
while weak ones ($\s_{ij}=-1$) have a synaptic weight $a-b$.
We assume $a$ and $b$ are constant all over the network.
All synapses are therefore excitatory for $b>0$, and inhibitory for $b<0$.

In the following, we focus our attention onto the slow plasticity dynamics
of the synaptic strength variables $\s_{ij}(t)$.
It will therefore be sufficient to consider the mean activities $\nub_i(t)$
and the mean input field $\hb_i(t)$,
defined by averaging over a time window which is large
w.r.t.~the characteristic time scale of neuron firings,
but short w.r.t.~that of synaptic dynamics.
These mean quantities obey
\beq
\nub_i(t)=F(\hb_i(t))
\eeq
and
\beq
\hb_i(t)=\frac{1}{N-1}\sum_{j\ne i}(a+b\s_{ij}(t))\nub_j(t).
\eeq

In most of this work we shall consider a spatially homogeneous situation
in the thermodynamic limit of a large network.
In this case the key quantity is the mean synaptic strength
\beq
J(t)=\frac{2}{N(N-1)}\sum_{(ij)}\s_{ij}(t),
\eeq
which does not fluctuate anymore.
The mean neuronal activity $\nub(t)$ and the mean input field $\hb(t)$
are related to $J(t)$ by the coupled non-linear equations
\beq
\nub(t)=F(\hb(t))
\label{nub}
\eeq
and
\beq
\hb(t)=(a+bJ(t))\nub(t).
\label{hb}
\eeq

Consider first the case where there are as many strong and weak synapses,
so that the mean synaptic strength vanishes ($J=0$).
We have then $\hb=a\nub$,
so that the mean neuronal activity $\nub$ obeys $\nub=F(a\nub)$.
We assume that the solution to that equation is $\nub=\frat{1}{2}$,
meaning that there are also as many active and inactive neurons on average.
We further simplify the problem by linearising
the coupled equations~(\ref{nub}),~(\ref{hb})
around this symmetric fixed point.
We thus obtain the following expression:
\beq
\nub(t)=f(J(t))=\frat{1}{2}(1+\eps J(t)).
\label{eff}
\eeq
The slope of the effective response function,
\beq
\eps=\frac{bF'(\frat{a}{2})}{1-aF'(\frat{a}{2})},
\eeq
is one of the key parameters of the model.\footnote{Here and
  throughout the following, primes denote derivatives.}
It has to obey $\abs{\eps}<1$.
It is positive in the excitatory case ($b>0$),
so that $f(J)$ is an increasing function of $J$,
and negative in the inhibitory case ($b<0$),
so that $f(J)$ is a decreasing function of $J$.

\subsubsection*{Synaptic plasticity dynamics}

Synaptic strengths evolve very slowly in time,
compared to the fast time scale of the firing rates of neurons.
It is therefore natural to model synaptic dynamics as a stochastic process
in continuous time~\cite{kampen},
defined in terms of effective jump rates between the two values
(strong or weak) of the synaptic strength.

The model includes the following three plasticity mechanisms
which drive synaptic evolution:

\begin{itemize}

\item[1.]{\it Spontaneous relaxation mechanism}.
  Synapses may spontaneously change their strength type,
  either from weak to strong (potentiation)
  or from strong to weak (depression)
  as a result of noise
  This spontaneous relaxation mechanism, illustrated in Figure~\ref{relax1},
  translates into
  \beq
  \begin{cases}
    \s_{ij}=-1\to+1&\mbox{with rate $\O$},\hfill\cr
    \s_{ij}=+1\to-1&\mbox{with rate $\o$}.\hfill\cr
  \end{cases}
  \eeq

  \begin{figure}[!ht]
    \begin{center}
      \includegraphics[angle=-90,width=.35\linewidth]{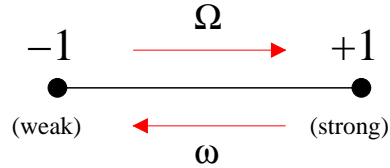}
      \caption{The spontaneous relaxation plasticity mechanism,
        with its potentiation rate $\O$ and depression rate $\o$
      (after Ref.~\cite{luck2014slow}).}
      \label{relax1}
    \end{center}
  \end{figure}

\item[2.]{\it Hebbian mechanism}.
  When two neurons are in the same state of (in)activity,
  the synapse which connects them strengthens;
  when one of the neurons is active and the other is not,
  the interconnecting synapse weakens.
  This is the well-known Hebbian mechanism~\cite{hebb},
  which we implement as follows:
  \beq
  \begin{cases}
    \nu_i(t)=\nu_j(t):\;\s_{ij}=-1\to+1&\mbox{with rate $\a$},\hfill\cr
    \nu_i(t)\ne\nu_j(t):\;\s_{ij}=+1\to-1&\mbox{with rate $\a$}.\hfill\cr
  \end{cases}
  \eeq

\item[3.]{\it Polarity mechanism}.
  This is a mechanism to introduce synaptic competition,
  introduced for the first time in~\cite{gmam1,gmam2},
  which converts a given synapse to the type of its most `successful' neighbours,
  i.e., those which augment the firing of an intermediate neuron.
  Thus: if a synapse $(ij)$ connects two neurons with different
  activities at time $t$, e.g.~$\nu_i(t)=+1$ and $\nu_j(t)=-1$,
  it will adapt its strength to that of a randomly selected synapse
  $(ik)$ connected to the active neuron $i$.
  If the selected synapse is strong,
  the update $\s_{ij}=-1\to+1$ takes place with rate $\b$;
  if it is weak,
  the update $\s_{ij}=+1\to-1$ takes place with rate $\g$.
  Therefore:
  \beq
  \begin{cases}
    \s_{ij}=-1\to+1&\mbox{with rate $\frat{1}{2}\b(1+J(t))$},\hfill\cr
    \s_{ij}=+1\to-1&\mbox{with rate $\frat{1}{2}\g(1-J(t))$}.\hfill\cr
  \end{cases}
  \eeq

\end{itemize}

\subsubsection*{Mean-field dynamics}\label{mf}

For a spatially homogeneous situation in the thermodynamic limit,
the mean synaptic strength $J(t)$
obeys a nonlinear dynamical mean-field equation of the form
\beq
\frac{\d J}{\d t}=P(J).
\label{djdt}
\eeq
The explicit form of the rate function $P(J)$ is obtained by summing the
contributions of the above three plasticity mechanisms.
%We thus obtain:
%\beq
%P(J)=P_1(J)+P_2(J)+P_3(J),
%\eeq
%with
%\beqa
%P_1(J)&=&\O(1-J)-\o(1+J),
%\nonumber\\
%P_2(J)&=&\a\left((2f(J)-1)^2-J\right)
%=-\a J(1-\eps^2J),
%\nonumber\\
%P_3(J)&=&-4\del(1-J^2)f(J)(1-f(J))
%=-\del(1-J^2)(1-\eps^2J^2),
%\label{p123}
%\eeqa
%where
%\beq
%\del=\frat{1}{4}(\g-\b).
%\eeq
In the most general situation, the model has five parameters:
the slope $\eps$ of the effective response function~(\ref{eff})
and the rates involved in the three plasticity mechanisms.
The resulting rate function is a polynomial of degree~4~\cite{luck2014slow}:
\beq
P(J)=p_4J^4+p_2J^2-(\O+\o+\a)J+\O-\o-\del,
\label{pj}
\eeq
with
\beq
p_4=-\del\eps^2,\qquad p_2=(\a+\del)\eps^2+\del,\qquad\del=\frat{1}{4}(\g-\b).
\label{p2p4}
\eeq
The spontaneous relaxation mechanism yields a linear rate function,
while the Hebbian mechanism is responsible for a quadratic non-linearity
and the polarity-driven competitive mechanism is responsible
for a quartic non-linearity.
This modelling of synaptic competition
satisfies the requirement on nonlinearity set out in Section~\ref{comp} for meaningful synaptic dynamics.

The parameter $\eps$ only enters~(\ref{p2p4}) through its square~$\eps^2$.
The model therefore exhibits an exact symmetry
between the excitatory case ($\eps>0$) and the inhibitory one ($\eps<0$).
(Since none of the plasticity mechanisms distinguishes between these two cases,
this symmetry is to be expected).
More generally, the model is invariant if the effective response function $f(J)$
is changed into $1-f(J)$.

\subsubsection*{Generic dynamics}

\begin{figure}[!ht]
  \begin{center}
    \includegraphics[angle=-90,width=.45\linewidth]{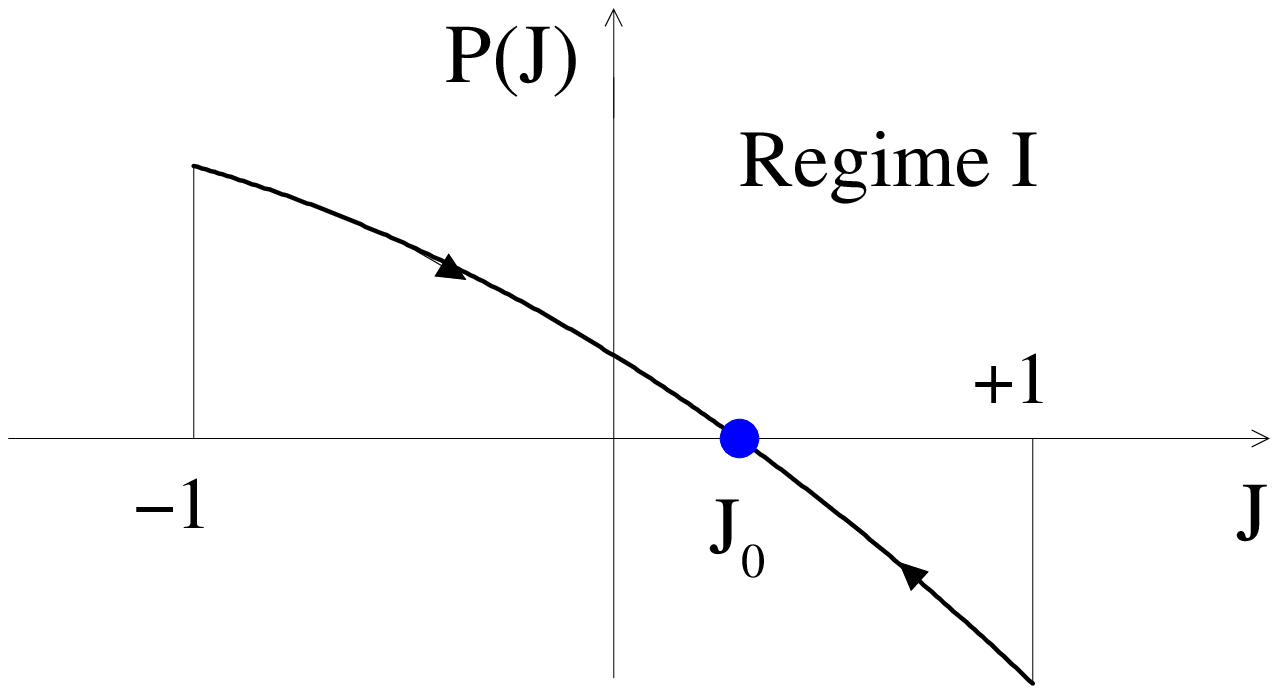}
    {\hskip 10pt}
    \includegraphics[angle=-90,width=.45\linewidth]{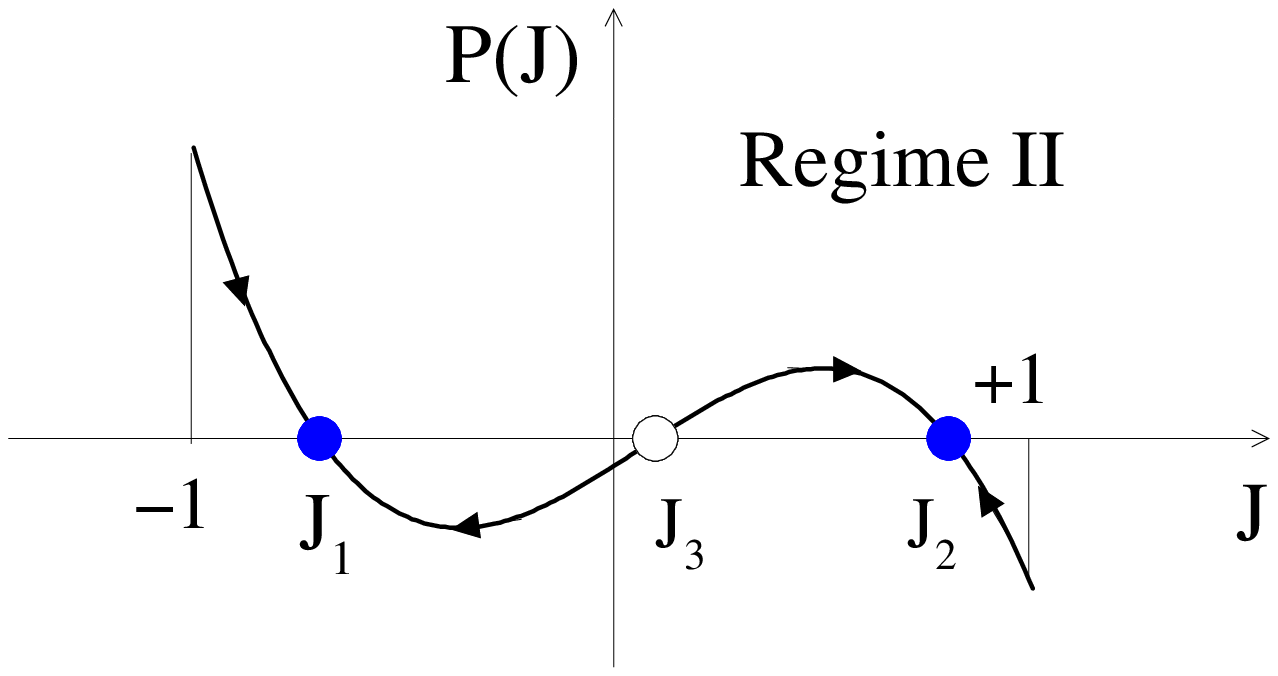}
    \caption{The two possible generic dynamical regimes.
      Left: Regime~I (one single attractive fixed point, $J_0$).
      Right: Regime~II (two attractive fixed points, $J_1$ and $J_2$,
      and an intermediate repulsive one, $J_3$)
      (after Ref.~\cite{luck2014slow}).}
    \label{regimes}
  \end{center}
\end{figure}

The rate function $P(J)$ has an odd number of zeros in the interval $-1<J<+1$
(counted with multiplicities), i.e., either one or three.
These zeros correspond to fixed points of the dynamics.
As a consequence, the model exhibits two generic dynamical regimes,
as shown in Figure~\ref{regimes}.

In Regime~I (see Figure~\ref{regimes}, left),
there is a single attractive (stable) fixed point at~$J_0$.
The mean synaptic strength $J(t)$ therefore converges exponentially fast
to this unique fixed point,
irrespective of its initial value, according to
\beq
J(t)-J_0\sim\e^{-t/\tau_0}.
\eeq
The corresponding relaxation time $\tau_0$ reads
\beq
\tau_0=-\frac{1}{P'(J_0)}.
\label{tau0}
\eeq
%In the limiting situation where there is only spontaneous relaxation,
%so that $P(J)=P_1(J)$, we have
%\beq
%J_0=\frac{\O-\o}{\O+\o},\qquad\tau_0=\frac{1}{\O+\o}.
%\eeq
where  $\tau_0$ and $J_0$ are obtainable in terms of the model parameters~\cite{luck2014slow}.

In Regime~II (see Figure~\ref{regimes}, right),
there are two attractive (stable) fixed points at $J_1$ and~$J_2$,
and an intermediate repulsive (unstable) one at $J_3$.
The mean synaptic strength $J(t)$ converges exponentially fast to
either of the attractive fixed points, depending on its initial value,
namely to $J_1$ if $-1<J(0)<J_3$ and to $J_2$ if $J_3<J(0)<+1$.
The corresponding relaxation times read
\beq
\tau_1=-\frac{1}{P'(J_1)},\qquad\tau_2=-\frac{1}{P'(J_2)}.
\label{tau12}
\eeq
In other words, Regime~II allows for the coexistence
of two separate fixed points,
leading to network configurations which are composed of largely strong/weak synapses.
In fact, it is the polarity-driven {\it competitive} mechanism which 
gives rise to the quartic non-linearity,
 essential for such coexistence.

\subsubsection*{Critical dynamics}

\begin{figure}[!ht]
  \begin{center}
    \includegraphics[angle=-90,width=.45\linewidth]{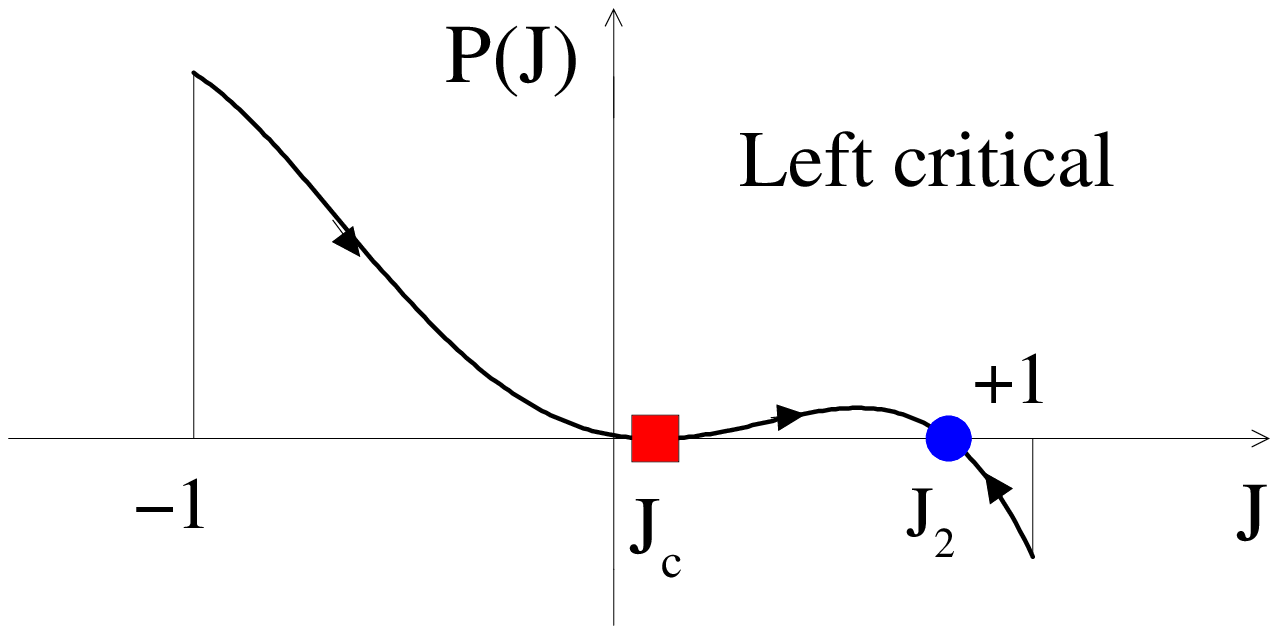}
    {\hskip 10pt}
    \includegraphics[angle=-90,width=.45\linewidth]{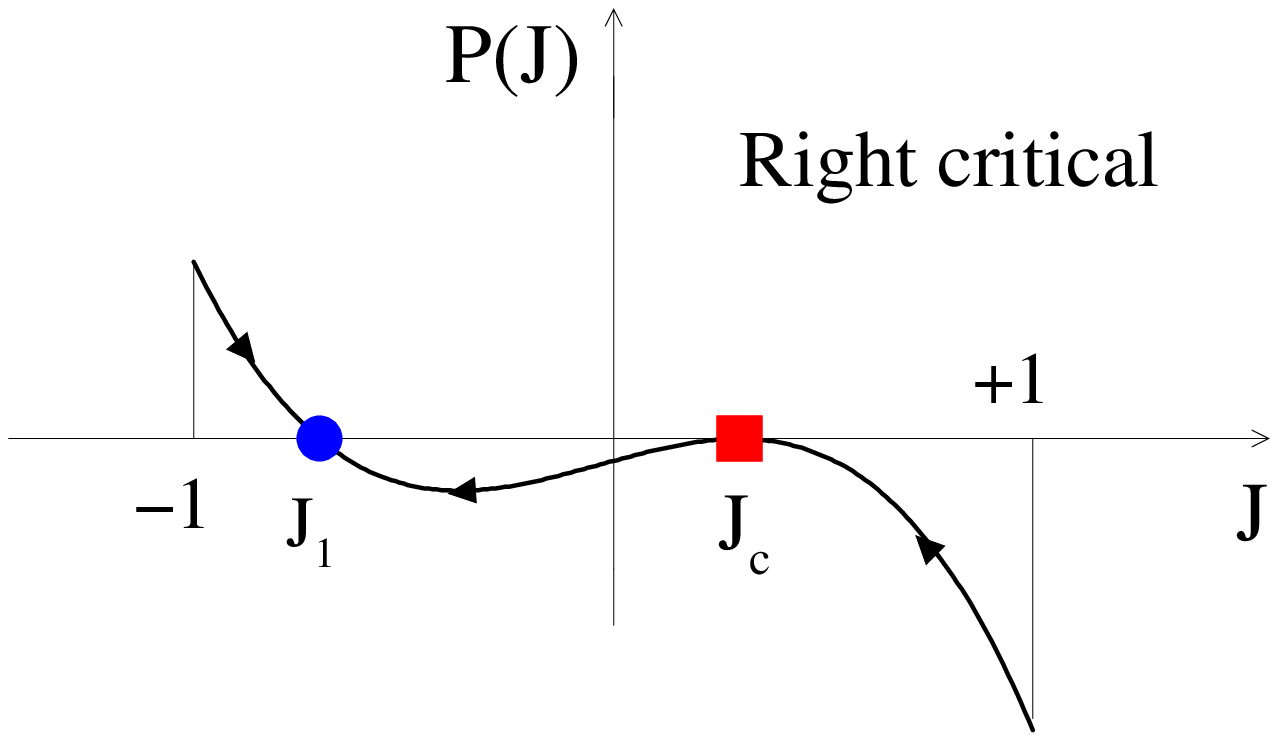}
    \caption{The two possible kinds of critical dynamical behaviour:
      left critical case ($J_1=J_3=J_c^{(L)}$)
      and right critical case ($J_2=J_3=J_c^{(R)}$)
      (after Ref.~\cite{luck2014slow}).}
    \label{critical}
  \end{center}
\end{figure}

When two of the three fixed points merge at some $J_c$,
the dynamical system~(\ref{djdt}) exhibits a saddle-node bifurcation.
In physical terms, the dynamics become critical.
We have then
\beq
P(J_c)=P'(J_c)=0,
\label{double}
\eeq
so that the critical synaptic strength $J_c$
is a double zero of the rate function $P(J)$ (see Figure~\ref{critical}).
There is a left critical case, where $J_1=J_3=J_c^{(L)}$, while $J_2$ remains non-critical,
and a right one, where $J_2=J_3=J_c^{(R)}$, while $J_1$ remains non-critical.
The critical synaptic strength obeys $J_c>\frat{1}{3}$~\cite{luck2014slow}.
We thus conclude that the critical point is always {\it strengthening},
as $J_c$ is always larger then the `natural' initial value $J(0)=0$,
corresponding to a random mixture of strong and weak synapses in equal proportions.

The mean synaptic strength exhibits a universal power-law relaxation
to its critical value, of the form
\beq
J(t)-J_c\approx\frac{A_c}{t}.
\label{tin}
\eeq
%The corresponding amplitude reads
%\beq
%A_c=-\frac{2}{P''(J_c)}=\frac{1}{6\del\eps^2(J_c^2-J_T^2)}.
%\eeq
%The formula~(\ref{jt2}) for $J_T^2$ has been used
%to derive the rightmost expression.
The asymptotic $1/t$ relaxation law~(\ref{tin})
holds irrespective of the initial value $J(0)$,
provided it is on the attractive side of the critical point,
i.e., $-1<J(0)<J_c$ in the left critical case (where $A_c<0$),
or $J_c<J(0)<+1$ in the right critical one (where $A_c>0$).

%\subsection{Tricritical dynamics}

%When all three fixed points merge at some $J_T$,
%the dynamical system~(\ref{djdt}) exhibits a pitchfork bifurcation.
%In physical terms, this corresponds to tricritical behaviour.
%We have then
%\beq
%P(J_T)=P'(J_T)=P''(J_T)=0,
%\label{triple}
%\eeq
%so that the tricritical synaptic strength $J_T$
%is a triple zero of the rate function $P(J)$ (see Figure~\ref{tricritical}).

%\begin{figure}[!ht]
%  \begin{center}
%    \includegraphics[angle=-90,width=.45\linewidth]{tricritical.eps}
%    \caption{\small
%      Tricritical dynamics.}
%    \label{tricritical}
%  \end{center}
%\end{figure}

%The mean synaptic strength again exhibits a universal power-law relaxation
%to its tricritical value $J_T$,
%albeit with a smaller exponent:
%\beq
%J(t)-J_T\approx\pm\frac{B_T}{\sqrt{t}}.
%\eeq
%This asymptotic $1/\sqrt{t}$ relaxation law holds irrespective of the initial value $J(0)$,
%with $\pm$ denoting the sign of the initial difference $J(0)-J_T$, whereas
%\beq
%B_T=\sqrt{-\frac{3}{P'''(J_T)}}=\frac{1}{\sqrt{8\del\eps^2J_T}}.
%\eeq
%The rightmost expression makes sense as $J_T$ is always positive.
%In other words, the tricritical point too is always strengthening.
%We have indeed $J_T>\frat{1}{3}$~(see~(\ref{tiers})).

To sum up, the non-critical fixed points of Regimes~I or~II
are characterised by exponential relaxation;
the corresponding relaxation times, whether long or short, are always finite.
Anywhere along the critical manifold, on the other hand,
one observes a universal power-law relaxation in $1/t$.
%An even slower power-law relaxation in $1/\sqrt{t}$ holds at the tricritical point.
Such behaviour corresponds to an infinite relaxation time at least in terms of the mean synaptic strength $J$.

In conclusion, this minimal model is able to show the emergence of power-law relaxation or long-term memory. It is clear that the most crucial
one of these is the mechanism of synaptic competition, which is in reassuring accord with the importance given to such competition by neuroscientists~\cite{avy} (Section~\ref{comp}).
Purely analytical work is able, however, just to give a flavour of the emergence of long-term behaviour in this model
via the critical behaviour of the mean synaptic strength $J$.
If realistic learning and forgetting of patterns are to be implemented with this model,
considerable computational work needs to be done. Only the identification of the parameter spaces where criticality is obtained in response to random input patterns will clarify, at least
phenomenologically, the routes to long-term memory in this relatively minimal model.

\section{Discussion}\label{disc}

Even quantitative approaches to the subject of memory are truly interdisciplinary; contributions range from mathematical psychology through
quantitative neuroscience to statistical physics. The narrowing of focus to physics still provides a huge range of contributions: from the seminal contributions
on Hopfield networks with their spin-glass analogies, through the emphasis on causality with spiking neurons, both of which involve fast neuronal dynamics, to the synaptic-dynamics-centred
approaches that have followed, with the boundedness of synaptic weights on discrete synapses, involving multiple `hidden' synaptic states, as well as the attribution of competitive and cooperative
dynamics to synapses in model networks. In this review, we have sought to highlight those approaches which generate {\it long-term memory}; while short-term memory, characterised by exponential relaxation times, is ubiquitous, long-term memory is characterised by power-law forgetting, a much slower process.

Another emphasis of this review is on synaptic competition, whose importance has long been understood by the neuroscience community, but which has only very recently been explicitly included
in model networks.This review has gone into as much detail in the need for this mechanism, as its inclusion in biophysical as well as physics-based modelling. In the latter case, the recent
advent of agent-based modelling techniques derived from game theory~\cite{vonn} and extended to cover nonequilibrium situations, has been particularly useful.

What is still a matter of debate is the extent to which phenomenological models, on which this review has focused, are useful in unravelling the phenomenon of memory storage and recall. While
it is certainly true that detailed biophysical models are overall better in matching experimental data point by point, there is a great deal to be said in favour of the formulation of minimal
models. These can, unlike the former, at least benefit from a few analytical insights, which can help both experimentalists and theorists identify the parameters that are truly important in what are 
typically huge parameter spaces, most recently believed to be in eleven dimensions~\cite{bluebrain}. While these large parameter spaces are indeed inclusive by definition, their inner workings can
only be described by computer simulations, which do not always give unambiguous answers to the relative importance of parameters, or answers to physical questions like, what are the crucial
mechanisms for memory storage? This is of course not to minimise their importance; we wish only to underscore the complementarity of the insights obtained by minimal physical models to the enigma
of memory.

\section*{Acknowledgements}

AM is very grateful to Dr. Jean-Marc Luck for a careful reading of this manuscript. She also
thanks the Institut f\"ur Informatik, Leipzig, and the University of Rome `La Sapienza' for their warm hospitality
during the course of this work.
This project has received funding from the European Research Council (ERC)
under the European Union’s Horizon 2020 Research and Innovation Programme
(grant agreement N.~694925).

\bibliographystyle{tADP}
\bibliography{anita}

\end{document}